\documentclass[reprint,amsmath,amssymb,aps,prl,showpacs,twocolumn]{revtex4}
\usepackage[utf8]{inputenc}
\usepackage{graphicx}
\usepackage{dcolumn}
\usepackage{bm}
\usepackage[breaklinks]{hyperref}
\usepackage{threeparttable}
\usepackage{bbold}
\usepackage{subfigure}
\usepackage{color}
\usepackage[normalem]{ulem}
\usepackage{cancel}
\usepackage{braket}
\usepackage{mathrsfs}

\renewcommand*{\HyperDestNameFilter}[1]{\jobname-#1}

\newcommand{\vn}[1]{{\boldsymbol{#1}}}

\begin{document}

\preprint{APS/123-QED}

\setcounter{secnumdepth}{2} 

\title{Spin-orbit torques and spin accumulation in FePt/Pt and Co/Cu thin films from first principles: the role of impurities}

\author{Guillaume G\'eranton}
\email{g.geranton@fz-juelich.de}
\author{Bernd Zimmermann}
\author{Nguyen Hoang Long}
\author{Phivos Mavropoulos}
\author{Stefan Bl\"ugel}
\author{Frank  Freimuth}
\email{f.freimuth@fz-juelich.de}
\author{Yuriy Mokrousov}
\email{y.mokrousov@fz-juelich.de}
\affiliation{Peter Gr\"unberg Institut and Institute for Advanced Simulation,
Forschungszentrum J\"ulich and JARA, 52425 J\"ulich, Germany}

\date{\today}

\begin{abstract}

Using the Boltzmann formalism based on the first principles electronic structure and scattering rates, we investigate the current-induced spin accumulation and spin-orbit torques in FePt/Pt and Co/Cu bilayers in the presence of substitutional impurities. In FePt/Pt bilayers we consider the effect
of intermixing of Fe and Pt atoms in the FePt layer, and find a crucial dependence of spin accumulation and spin-orbit torques on the details of the distribution of these defects. In Co/Cu bilayers we predict that the magnitude and sign of the spin-orbit torque and spin accumulation depend very sensitively on the type of the impurities used to dope the Cu substrate. Moreover, simultaneously with impurity-driven scattering we consider the effect of an additional constant quasiparticle broadening of the states at the Fermi surface to simulate phonon scattering at room temperature, and discover that even a small broadening of the order of 25\,meV can drastically influence the magnitude of the considered effects. We explain our findings based on the analysis of the complex interplay of several competing Fermi surface contributions to the spin accumulation and spin-orbit torques in
 these structurally and chemically non-uniform systems. 
\end{abstract}

\maketitle

\section{Introduction}

Spin-orbit torques (SOTs) rely on the spin-orbit mediated exchange of angular momentum between the crystal lattice and the magnetization in the presence of an electric field~\cite{PhysRevB.79.094422,PhysRevB.80.134403}. It was recently found that they are able  to switch the magnetization in ferromagnetic bilayers~\cite{Miron:155001,Liu04052012,PhysRevLett.109.096602}, and have attracted considerable interest for technological applications in the field of magnetic random access memories. Two different mechanisms have been suggested that give rise to SOTs in bilayers consisting of a heavy metal substrate and a thin ferromagnetic layer deposited on top of it. The first mechanism is attributed to the spin Hall effect \cite{Kato10122004,RevModPhys.87.1213} which generates a spin current from the subtrate towards the ferromagnet~\cite{Liu04052012,PhysRevLett.109.096602,PhysRevB.87.174411}. The second mechanism is due to the generation of a current-induced spin accumulation \cite{V.M._Edelstein_1990,MihaiMiron:155006,Miron:154509,PhysRevB.85.180404,PhysRevB.87.174411} at the interface between the two materials, where magnetism, spin-orbit coupling and broken inversion symmetry coexist. While the spin Hall conductivity of the heavy metal is a rather robust quantity, the current-induced spin accumulation generally depends very sensitively on the details of disorder at the interface. 

On the side of material-specific theory of spin-orbit torque, most \textit{ab initio} calculations of SOTs in ferromagnetic bilayers in the last few years were performed within the constant relaxation time approximation~\cite{PhysRevB.88.214417,ibcsoit,PhysRevB.92.064415,PhysRevB.91.014417}. While the importance of impurity scattering is well established in the field of relativistic charge and spin 
transport~\cite{PhysRevLett.105.266604,PhysRevLett.106.126601,PhysRevLett.109.156602,PhysRevLett.104.186403,PhysRevB.81.245109,PhysRevLett.106.056601,PhysRevB.92.184415}, the interplay between different types of disorder and current-induced SOTs in ferromagnetic heterostructures is essentially unexplored. Nonetheless, the crucial role of surfaces and interfaces giving rise to the spin accumulation suggests an enhanced sensitivity of the spin-orbit torque to structural and chemical types of disorder in these regions. This line of thought is supported by the large effect of annealing on the SOTs in AlO$_x$/Co/Pt and MgO/CoFeB/Ta thin films \cite{symmetry_spin_orbit_torques,PhysRevB.89.214419}. This calls for a
first principles theory of SOT that is able to account for an effect of specific 
types of defects and impurities, especially close to interfaces.

In this paper, we present an implementation of \textit{ab initio} Boltzmann formalism for the spin-orbit torque based on the Korringa-Kohn-Rostoker (KKR) Green functions method, which is ideally suited for studying the effect of impurity scattering on the SOT and spin accumulation. Within this formalism we first find the states on the Fermi surface (FS) by solving the KKR secular equation, and compute the corresponding expectation values of the velocity and torque operators. Next, we compute the scattering amplitudes of the states off single impurities and obtain in the dilute limit the corresponding transition rates and relaxation times for a finite concentration of impurities. Finally, after solving the linearized Boltzmann equation and determining the non-equilibrium distribution function of the system in the presence of an electric field and
impurities, we compute the spin accumulation and the SOTs as FS integrals
of corresponding matrix elements.

We apply the developed Boltzmann methodology to the investigation of current-induced SOTs in FePt/Pt and Co/Cu bilayers in the presence of substitutional impurities. In FePt/Pt bilayers we consider the influence of intermixing of Fe and Pt atoms in the FePt layer, and find a crucial dependence of the SOT on the distribution of defects. Moreover, we demonstrate that a large part of the SOT is mediated by spin currents and observe a large spin accumulation in the Pt layers. In Co/Cu bilayers we investigate the effect of doping with Bi, Ir, C and N impurities and find the magnitude and the sign of the SOT to be very sensitive to the type of impurities, which we explain by the state-dependent relaxation-time induced by scattering off defects. Finally, simultaneously with impurity-driven scattering we consider the effect of a constant smearing of the states to include other sources of scattering that exist at room temperature, such as phonons, and find that even a small smearing of the order of 25 meV can modify the SOT significantly.
 
The article is structured as follows. First, in Section~\ref{lFormalism}.A we present the KKR implementation of the torque, spin accumulation and spin-flux operators. Next, in Section~\ref{lFormalism}.B we provide the expressions for the transition rates, torques and spin accumulation within the Boltzmann formalism, and then in Section~\ref{lFormalism}.C we describe how to include effectively into consideration other sources of scattering present at room temperature. The results and discussion of our calculations of SOTs and 
spin accumulation in FePt/Pt and Co/Cu bilayers are presented in Section~\ref{Sec_results}. We conclude our study in Section~\ref{Conclusions}.

\section{Formalism}\label{lFormalism}
\subsection{KKR method} \label{ssec_KKRformalism}

We compute the electronic structure of the films within the local density approximation to density functional theory using the relativistic full-potential KKR method~\cite{0022-3719-20-16-010}. We find the states $\psi_{\vn{k}}$ on the Fermi surface by solving the KKR secular equation~\cite{Bernd_FS}. For each state on the Fermi surface we compute the spin expectation value. The contribution of a state to the $i$-th component of the spin at the atom $\mu$ is given by:
\begin{equation}\label{Sigma_operator}
%\begin{aligned}
%\langle\sigma_{i\mu}\rangle &= \langle\psi|\sigma_{i\mu}|\psi\rangle\\
%                                 &= \int_{V_{\mu}} d^{3}r~[\psi(\vn{r})]^{\dagger}\sigma_{i}\psi(\vn{r}).
%\end{aligned}
\langle\sigma_{i\mu}\rangle_{\vn{k}} = \langle\psi_{\vn{k}}|\sigma_{i\mu}|\psi_{\vn{k}}\rangle
                                 = \int_{V_{\mu}} d^{3}r~[\psi_{\vn{k}}(\vn{r})]^{\dagger}\sigma_{i}\psi_{\vn{k}}(\vn{r}),
\end{equation}
where $\sigma_{i}$ is the $i$-th Pauli matrix and the integration is performed over the volume of the atomic cell $V_{\mu}$. In the KKR formalism, the wavefunction in the atomic cell $\mu$ with center $\vn{R}_{\mu}$ is expanded as:
\begin{equation}\label{KKRrep_WF}
\psi_{\vn{k}}(\vn{r}+\vn{R}_{\mu})=\sum_{L,s}c^{s,\mu}_{L}(\vn{k},\varepsilon)R^{s,\mu}_{L}(\vn{r},\varepsilon),
\end{equation}
where $R^{s,\mu}_{L}$ are the regular solutions of the single-site scattering problem for site $\mu$ at the energy $\varepsilon$ and $c^{s,\mu}_{L}$ are the expansion coefficients. The spin index $s$ and the combined index $L=\{l,m\}$ for azimuthal quantum number $l$ and magnetic quantum number $m$ denote the incoming boundary conditions. In the presence of spin-orbit coupling, both $\psi$ and $R^{s,\mu}_{L}$ are 4-vectors that allow for a component of the spin perpendicular to the spin quantization axis. The spin polarization of a state $\psi$ on site $\mu$, which gives the contribution of the state to the spin accumulation, reads in the KKR representation:
\begin{equation}\label{KKRrep_sa}
\langle\sigma_{i\mu}\rangle_{\vn{k}}=\sum_{L s}\sum_{L^{\prime} s^{\prime}}
[c^{s,\mu}_{L}]^*\,\Sigma^{s s^{\prime},\mu}_{L L^{\prime},i}\,c^{s^{\prime},\mu}_{L^{\prime}}.
\end{equation}
The matrix elements $\Sigma^{s s^{\prime},\mu}_{L L^{\prime},i}$ have already been given in~\cite{Heers:82648,Bernd_FS}.

The torque exerted by the exchange field on the electronic states is given by the vector product of the spin magnetic moment $-\mu_{B}\boldsymbol{\sigma}$ with the exchange field $\vn{B}(r)$, which we compute in this work from first principles for a given system. The components of the torque operator are given by:
\begin{equation}\label{KKRrep_Tdef}
\mathcal{T}_{i}(\vn{r})=-\mu_{B}\sum_{j k}\epsilon_{ijk}\sigma_{j}B_{k}(\vn{r}),
\end{equation}
where $\epsilon_{ijk}$ is the Levi-Civita symbol and the indices $i$, $j$ and $k$ are $x$, $y$ or $z$. The torque exerted on a state at the atom $\mu$ is given by:
\begin{equation}\label{T_operator}
\begin{aligned}
\langle\mathcal{T}_{i\mu}\rangle_{\vn{k}} &= \langle\psi_{\vn{k}}|\mathcal{T}_{i\mu}|\psi_{\vn{k}}\rangle\\
                                 &= -\mu_{B}\sum_{j k}\epsilon_{ijk}\int_{V_{\mu}} d^{3}r~[\psi_{\vn{k}}(\vn{r})]^{\dagger}\sigma_{j}\psi_{\vn{k}}(\vn{r})B_{k}(\vn{r}).
\end{aligned}
\end{equation}
According to Eq.~\eqref{KKRrep_WF}, the torque expectation values in KKR thus take the form:
\begin{equation}\label{KKRrep_T}
\langle\mathcal{T}_{i\mu}\rangle_{\vn{k}}=\sum_{L s}\sum_{L^{\prime} s^{\prime}}
[c^{s,\mu}_{L}]^*\,\mathfrak{T}^{s s^{\prime},\mu}_{L L^{\prime},i}\,c^{s^{\prime},\mu}_{L^{\prime}},
\end{equation}
and the matrix elements $\mathfrak{T}^{s s^{\prime},\mu}_{L L^{\prime},i}$ are defined by
\begin{equation}
\begin{aligned}
\mathfrak{T}^{s s^{\prime},\mu}_{L L^{\prime},i} = &-\mu_{B}\sum_{j k}\epsilon_{ijk}\\
                                                   &\int d^{3}r~\theta^{\mu}(\vn{r})[R^{s,\mu}_{L}(\vn{r},\varepsilon_{\rm F})]^{\dagger}\sigma_{j}R^{s^{\prime},\mu}_{L^{\prime}}(\vn{r},\varepsilon_{\rm F})B^{\mu}_{k}(\vn{r}).
\end{aligned}
\end{equation}
%\begin{equation}\label{Top_1}
%\begin{aligned}
%\mathfrak{T}^{s s^{\prime},\mu}_{L L^{\prime},i} = &-\mu_{B}\sum_{j k}\epsilon_{ijk}\\
%                                                   &\int_{V_{\mu}} d^{3}r~[R^{s,\mu}_{L}(\vn{r},\varepsilon_{\rm F})]^{\dagger}\sigma_{j}R^{s^{\prime},\mu}_{L^{\prime}}(\vn{r},\varepsilon_{\rm F})B^{\mu}_{k}(\vn{r}),
%\end{aligned}
%\end{equation}
where we used the notation $B^{\mu}_{k}(\vn{r})=B_{k}(\vn{r}+\vn{R}_{\mu})$. The shape functions $\theta^{\mu}(\vn{r})$ \cite{STEFANOU1990231, 0953-8984-3-39-006} of the Voronoi cells allows us to extend the integration to the entire space. In the following we always consider the regular scattering solutions $R^{s,\mu}_{L}(\vn{r},\varepsilon_{\rm F})$ at the Fermi energy and therefore omit the energy index.

The scattering solutions $R^{s,\mu}_{L}(\vn{r})$, the shape functions $\theta^{\mu}(\vn{r})$ and the exchange field $B^{\mu}_{k}(\vn{r})$ are all expanded in real spherical harmonics as:
\begin{eqnarray}\label{KKRrep_WF_scatsol}
R^{s,\mu}_{L}(\vn{r})&=&\sum_{L_{2}}\frac{1}{r} R^{s,\mu}_{L_{2} L}(r) Y_{L_{2}}(\hat{\vn{r}}),\\
\theta^{\mu}(\vn{r})&=&\sum_{L_{1}}\theta^{\mu}_{L_{1}}(r) Y_{L_{1}}(\hat{\vn{r}}),\\
B^{\mu}_{k}(\vn{r})&=&\sum_{L_{4}}B^{\mu}_{L_{4},k}(r) Y_{L_{4}}(\hat{\vn{r}}),
\end{eqnarray}
where $r=|\vn{r}|$ and $\hat{\vn{r}}=\vn{r}/|\vn{r}|$ are respectively the magnitude and the direction of $\vn{r}$. To avoid the integration of the product of four spherical harmonics, we first compute the convoluted exchange field $b^{\mu}_{k}(\vn{r})=B^{\mu}_{k}(\vn{r})\theta^{\mu}(\vn{r})$ and then replace the two spherical harmonics expansions of $B^{\mu}_{k}(\vn{r})$ and $\theta^{\mu}(\vn{r})$ by a single one for $b^{\mu}_{k}(\vn{r})$:
\begin{equation}
b^{\mu}_{k}(\vn{r})=\sum_{L_{5}}b^{\mu}_{L_{5},k}(r) Y_{L_{5}}(\hat{\vn{r}}).
\end{equation}
The matrix elements of the torque operator finally read:
\begin{equation}\label{Top_3}
\begin{aligned}
\mathfrak{T}^{s s^{\prime},\mu}_{L L^{\prime},i} = &-\mu_{B}\sum_{j k}\epsilon_{ijk}\sum_{L_2 L_3 L_5}C_{L_2 L_3 L_5}\\
                                                   &\int d^{3}r~[R^{s,\mu}_{L_2 L}(r)]^{\dagger}\sigma_{j}R^{s^{\prime},\mu}_{L_3 L^{\prime}}(r)b^{\mu}_{L_{5},k}(r),
\end{aligned}
\end{equation}
%\begin{equation}\label{Top_2}
%\begin{aligned}
%\mathfrak{T}^{s s^{\prime},\mu}_{L L^{\prime},i} = &-\mu_{B}\sum_{j k}\epsilon_{ijk}\sum_{L_1 L_2 L_3 L_4}\\
%\times &\int d\Omega~Y_{L_{1}}(\hat{\vn{r}})Y_{L_{2}}(\hat{\vn{r}})Y_{L_{3}}(\hat{\vn{r}})Y_{L_{4}}(\hat{\vn{r}})\\
%\times &\int dr~\theta^{\mu}_{L_1}(r)[R^{s,\mu}_{L_2 L}(r)]^{\dagger}\sigma_{j}R^{s^{\prime},\mu}_{L_3 L^{\prime}}(r)B^{\mu}_{L_{4},k}(r),
%\end{aligned}
%\end{equation}
%\begin{equation}
%\begin{aligned}
%b^{\mu}_{L_{5},k}(r)&=\sum_{L_{1}L_{4}}\int %d\Omega~Y_{L_{1}}(\hat{\vn{r}})Y_{L_{4}}(\hat{\vn{r}})Y_{L_{5}}(\hat{\vn{r}})\theta^{\mu}_{L_1}(r)B^{\mu}_{L_{4},k}(r)\\
%                                        &=\sum_{L_{1}L_{4}}C_{L_1 L_4 L_5}\theta^{\mu}_{L_1}(r)B^{\mu}_{L_{4},k}(r),
%\end{aligned}
%\end{equation}
where $C_{L_1 L_4 L_5}$ are the Gaunt coefficients.

The torque in a ferromagnet/heavy metal bilayer usually has a large contribution arising from spin currents \cite{ibcsoit}. It is therefore very instructive to compare the atom-resolved torques to the spin fluxes into the corresponding atoms. We derive in the following the expression for the spin flux operator in KKR. The contribution of a state to the spin flux flowing into the atom $\mu$ is given by:
\begin{equation}
\begin{aligned}
\langle\mathcal{Q}_{i\mu}\rangle_{\vn{k}}&=-\frac{\mu_{B}\hbar}{2ie}\int_{S_{\mu}}d\vn{S}\cdot[\psi_{\vn{k}}^{\dagger}(\vn{r})\sigma_{i}\boldsymbol{\nabla}\psi_{\vn{k}}(\vn{r})\\
                                          &\phantom{=-\frac{\mu_{B}\hbar}{2ie}\int_{S_{\mu}}d}-\boldsymbol{\nabla}\psi_{\vn{k}}^{\dagger}(\vn{r})\sigma_{i}\psi_{\vn{k}}(\vn{r})],
\end{aligned}
\end{equation}
where the surface $S_{\mu}$ corresponds to the muffin-tin (MT) sphere of the atom $\mu$. In the KKR representation the expectation values of the spin-flux operator read
\begin{equation}\label{KKRrep_Q}
\langle\mathcal{Q}_{i\mu}\rangle_{\vn{k}}=\sum_{L s}\sum_{L^{\prime} s^{\prime}}
[c^{s,\mu}_{L}]^*\,q^{s s^{\prime},\mu}_{L L^{\prime},i}\,c^{s^{\prime},\mu}_{L^{\prime}},
\end{equation}
where the matrix elements $q^{s s^{\prime},\mu}_{L L^{\prime},i}$ are defined by
\begin{equation}\label{KKRrep_qint}
\begin{aligned}
q^{s s^{\prime},\mu}_{L L^{\prime},i}=-\frac{\mu_{B} \hbar}{2ie} \int_{S_{\mu}}d\vn{S}&\cdot\Big[ [R^{s ,\mu}_{L}(\vn{r})]^{\dagger}\sigma_{i}\boldsymbol{\nabla}R^{s^{\prime} ,\mu}_{L^{\prime}}(\vn{r})\\
&- [\boldsymbol{\nabla}R^{s ,\mu}_{L}(\vn{r})]^{\dagger}\sigma_{i}R^{s^{\prime} ,\mu}_{L^{\prime}}(\vn{r})\Big].
\end{aligned}
\end{equation}
Taking into account that the integration takes place on the MT sphere, the above expression reduces to:
\begin{widetext}
\begin{equation}
\begin{aligned}\label{KKRrep_qint_3}
q^{s s^{\prime},\mu}_{L L^{\prime},i}= -\frac{\mu_{B} \hbar}{2ie} \sum_{L_{1}} \Big[
[R^{s ,\mu}_{L_{1}L}(r_{\rm MT})]^{\dagger}\sigma_{i} \frac{\partial}{\partial r}\Big(R^{s^{\prime} ,\mu}_{L_{1}L^{\prime}}(r)\Big)-R^{s^{\prime} ,\mu}_{L_{1}L^{\prime}}(r_{\rm MT})\sigma_{i} \frac{\partial}{\partial r}\Big( [R^{s ,\mu}_{L_{1}L}(r)]^{\dagger}\Big)\Big]_{r=r_{\rm MT}}.
\end{aligned}
\end{equation}
\end{widetext}

\subsection{Boltzmann equation}

We compute the response of the system to an external electric field within the Boltzmann formalism, in which the trajectory of the electrons between collisions obeys semiclassical equations of motion but the transition rates induced by the scattering off impurities are computed quantum mechanically. Let us first define the deviation of the distribution function $g_{n}(\vn{k})=f_{n}(\vn{k})-f^{0}_{n}(\varepsilon(\vn{k}))$ as the difference between the non-equilibrium distribution function $f_{n}(\vn{k})$ and the equilibrium Fermi-Dirac distribution $f^{0}_{n}(\varepsilon(\vn{k}))$, where $n$ is the band index and $\vn{k}$ the wave vector of the state. The Boltzmann equation for a homogeneous system of semiclassical electrons is 
\begin{equation}\label{eq_Boltzmann}
\dot{\vn{k}}\nabla_{\vn{k}}f_{n}(\vn{k}) = \sum_{\vn{k'}n'} \bigg(g_{n'}(\vn{k'}) P^{nn'}_{\vn{k}\vn{k'}} - g_{n}(\vn{k}) P^{n'n}_{\vn{k'}\vn{k}}\bigg),
\end{equation}
where $P^{nn'}_{\vn{k}\vn{k'}}$ is the rate of the transition from the state $n'\vn{k'}$ into the state $n\vn{k}$ that is induced by the scattering off impurities. The trajectory of the electron in $\vn{k}$-space when an electric field is applied is described by the semiclassical equation of motion $\hbar\dot{\vn{k}}=-e\vn{E}$, where $e=|e|$ is the positive elementary charge. For a weak electric field, the variation of the distribution function $g_{n}(\vn{k})$ is expected to be linear in $\vn{E}$ and sizeable only close to the Fermi energy, which suggests the following Ansatz:
\begin{equation}\label{eq_Ansatz_vmfp}
g_{n}(\vn{k})=e\frac{\partial f^{0}(\mathcal{E}_{n}(\vn{k}))}{\partial \mathcal{E}_{n}(\vn{k})} \mbox{{\boldmath$\lambda$}}_{n}(\vn{k}) \cdot \vn{E}.
\end{equation}
Inserting Eq.~\eqref{eq_Ansatz_vmfp} into Eq.~\eqref{eq_Boltzmann} yields a self-consistent equation for the vector mean free path $\mbox{{\boldmath$\lambda$}}$:
\begin{equation}\label{BOLTZ_Iter}
\mbox{{\boldmath$\lambda$}}(\vn{k}) = \tau_{\vn{k}}\bigg(\vn{v}(\vn{k}) + \sum_{\vn{k'}}P_{\vn{k}\vn{k'}}\mbox{{\boldmath$\lambda$}}(\vn{k'})\bigg),
\end{equation}
where $\vn{v}(\vn{k})$ is the group velocity and $\tau_{\vn{k}}$ is the relaxation time for state $\vn{k}$.  We drop the band indices in Eq.~\eqref{BOLTZ_Iter} because the transition rates $P_{\vn{k}\vn{k'}}$ are energy-conserving. The expression for the transition rates for a given kind of impurity can be found in Ref.~\cite{PhysRevLett.104.186403}. When different types of impurities are present (different chemical elements or different atomic sites), the total transition rates are given by 
\begin{equation}\label{transitionrates_gen}
P_{\vn{k}\vn{k'}}=\frac{2\pi}{\hbar}N\delta(\varepsilon(\vn{k})-\varepsilon(\vn{k'}))\sum_{m} c_{m} |T^{m}_{\vn{k}\vn{k'}}|^2 ,
\end{equation}
where $N$ is the number of unit cells in the sample, $c_{m}$ is the concentration of the impurities of type $m$ and $T^{m}_{\vn{k}\vn{k'}}$ are the elements of the $T$-matrix for impurities of type $m$ as defined in Ref.~\cite{PhysRevB.90.064406}. An expression similar to Eq.~\eqref{transitionrates_gen} has been used successfully in the past for the study of the spin Hall and spin Nernst effects in ternary alloys \cite{PhysRevB.87.161114}.
The relaxation times $\tau_{\vn{k}}$ are defined by:
\begin{equation}\label{relaxation_time}
\tau_{\vn{k}}^{-1}=\sum_{\vn{k'}}P_{\vn{k'}\vn{k}}.
\end{equation}
The knowledge of the vector mean free path $\mbox{{\boldmath$\lambda$}}(\vn{k})$ allows us to compute the spin accumulation $\vn{s}_{\mu}$ in terms of the induced magnetic moment, the torque $\vn{T}_{\mu}$ and the spin flux $\vn{Q}_{\mu}$ of the atom $\mu$ induced by an applied electric field. We define the response tensors for the spin accumulation, $\bm{\chi}_{\mu}$, the torque, $\vn{t}_{\mu}$, and the spin flux, $\vn{q}_{\mu}$, according to:
\begin{equation}\label{BOLTZ_safin3_3D}
\vn{s}_{\mu} = \bm{\chi}_{\mu}\vn{E},
\end{equation}
\begin{equation}\label{BOLTZ_Tfin3_3D}
\vn{T}_{\mu} = \vn{t}_{\mu}\vn{E},
\end{equation}
\begin{equation}\label{BOLTZ_qfin3_3D}
\vn{Q}_{\mu} = \vn{q}_{\mu}\vn{E}.
\end{equation}
All three response tensors take the form of Fermi surface integrals:
\begin{equation}\label{BOLTZ_sa_2D}
\bm{\chi}_{\mu} = \frac{e \mu_{B}}{\hbar \mathcal{S_{BZ}}} \int_{\rm FS} \frac{d\vn{k}}{|\vn{v}(\vn{k})|}~\langle \bm{\sigma}_{\mu} \rangle_{\vn{k}} \otimes \mbox{{\boldmath$\lambda$}}(\vn{k}),
\end{equation}
\begin{equation}\label{BOLTZ_t_2D_atom}
\vn{t}_{\mu} = \frac{e}{\hbar \mathcal{S_{BZ}}} \int_{\rm FS} \frac{d\vn{k}}{|\vn{v}(\vn{k})|}~\langle \bm{\mathcal{T}}_{\mu} \rangle_{\vn{k}} \otimes  \mbox{{\boldmath$\lambda$}}(\vn{k}),
\end{equation}
\begin{equation}\label{BOLTZ_q_2D}
\vn{q}_{\mu} = -\frac{e}{\hbar \mathcal{S_{BZ}}} \int_{\rm FS} \frac{d\vn{k}}{|\vn{v}(\vn{k})|}~\langle \bm{\mathcal{Q}}_{\mu} \rangle_{\vn{k}} \otimes  \mbox{{\boldmath$\lambda$}}(\vn{k}).
\end{equation}
The spin accumulation $\langle \bm{\sigma}_{\mu} \rangle_{\vn{k}}$, the torque $\langle \bm{\mathcal{T}}_{\mu} \rangle_{\vn{k}}$ and the spin flux $\langle \bm{\mathcal{Q}}_{\mu} \rangle_{\vn{k}}$ for the state $\psi_{\vn{k}}$ are computed according to Eqs.~\eqref{KKRrep_sa}, \eqref{KKRrep_T} and \eqref{KKRrep_Q}. Since all atoms in the unit cell are exchange coupled, it is relevant to consider the total torque $\vn{T}$ exerted on the magnetic moment of a unit cell. We define the total torkance $\vn{t}$ for the total torque according to:
\begin{equation}\label{BOLTZ_Tfin3_3D}
\vn{T} = \vn{t}\vn{E}.
\end{equation}
The total torkance reads
\begin{equation}\label{BOLTZ_t_2D}
\vn{t} = \frac{e}{\hbar \mathcal{S_{BZ}}} \int_{FS} \frac{d\vn{k}}{|\vn{v}(\vn{k})|}~\langle \bm{\mathcal{T}} \rangle_{\vn{k}} \otimes  \mbox{{\boldmath$\lambda$}}(\vn{k}),
\end{equation}
where $\bm{\mathcal{T}}=\sum_{\mu}\bm{\mathcal{T}}_{\mu}$ is the total torque operator.

\subsection{Finite temperature} \label{sec_finite_temp}

The transition rates computed from Eq.~\eqref{transitionrates_gen} capture the effect of the scattering off impurities. They are suitable to calculate residual resistivities and zero-temperature spin-orbit torques (0K-SOT). At room temperature, the scattering off phonons is crucial to explain conductivities in metals. In order to include the effect of a finite temperature, we add for all states a constant contribution to lifetimes and scattering probabilities.
We thus replace the lifetimes in Eq.~\eqref{BOLTZ_Iter} by $\widetilde{\tau}_{\vn{k}}$, which we define by the relation 
\begin{equation}\label{relaxtilde}
\frac{1}{\widetilde{\tau}_{\vn{k}}} = \frac{1}{\tau_{\vn{k}}} + \frac{2 \Gamma}{\hbar},
\end{equation}
where we set the parameter $\Gamma$ to 25 meV to simulate the effect of phonon scattering.
The matrix elements $P_{\vn{k}\vn{k'}}$ of the scattering-in term in Eq.~\eqref{BOLTZ_Iter} must be modified in a consistent way, i.e., it must be ensured that $\widetilde{\tau}_{\vn{k}}^{-1} = \sum_{\vn{k}'} \widetilde{P}_{\vn{k}'\vn{k}}$. This condition is fulfilled by introducing the generalized transition rates:
\begin{equation}\label{Pkk_tilde}
\widetilde{P}_{\vn{k}'\vn{k}} = P_{\vn{k}'\vn{k}} + \frac{2 \Gamma}{\hbar n(\varepsilon(\vn{k'}))}\delta(\varepsilon(\vn{k})-\varepsilon(\vn{k'})),
\end{equation}
where $n(\varepsilon)$ is the density of states.
%\begin{equation}
%n(\varepsilon)=\sum_{\vn{k}}\delta(\varepsilon(\vn{k})-\varepsilon).
%\end{equation}
The relaxation times and the transition rates from Eq.~\eqref{relaxtilde} and \eqref{Pkk_tilde} can be used to compute the effect of specific types of defects or impurities in a system where other sources of scattering exist, as it is the case at room temperature (RT).

\section{Results}\label{Sec_results}

We now apply the formalism described in Section \ref{lFormalism} to compute the SOT in thin films where impurities and defects are present. We model different types of disorder by considering different profiles $c_{m}$ of impurity concentration, where $m$ is the layer index. Thus, $c_{m}$ gives the average number of impurities per unit cell in a given atomic layer, and $\bar{c}_{\rm imp}=\sum_{m}c_{m}$ is the average number of impurities per unit cell for all layers. The $c_{m}$ coefficients thereby defined are used in Eq.~\eqref{transitionrates_gen} along with the corresponding $T^{m}_{\vn{k}\vn{k'}}$ matrix elements to compute the transition rate $P_{\vn{k}\vn{k'}}$ for each impurity profile. The transition rates are then used to compute the vector mean free path $\mbox{{\boldmath$\lambda$}}(\vn{k})$ from Eq.~\eqref{BOLTZ_Iter} and the zero-temperature response tensors for the spin accumulation $\bm{\chi}^{\rm 0K}_{\mu}$, the torque $\vn{t}^{\rm 0K}_{\mu}$ and the spin fluxes $\vn{q}^{\rm 0K}_{\mu}$. In order to account for other sources of scattering that exist at room temperature, we also compute the generalized transition rates using Eq.~\eqref{Pkk_tilde} instead of Eq.~\eqref{transitionrates_gen}, which allows the calculation of the room temperature response tensors for the spin accumulation $\bm{\chi}^{\rm RT}_{\mu}$, the torque $\vn{t}^{\rm RT}_{\mu}$ and the spin fluxes $\vn{q}^{\rm RT}_{\mu}$.

Because SOTs have been investigated in the past with the Kubo formalism in the constant relaxation time approximation (CRTA) \cite{ibcsoit,PhysRevB.91.014417}, we want to clarify the relation to the Boltzmann formalism we are using here. In the absence of impurities and at finite $\tau$, the torkance we compute reduces to the intraband terms of the torkance in Ref.~\cite{PhysRevB.91.014417}, which is always an odd function of magnetization direction. In the presence of impurities, the relaxation time $\tau_{\vn{k}}$ becomes state dependent, see Eq.~\eqref{relaxation_time}. This state dependent relaxation time enters the first term on the right hand side of Eq.~\eqref{BOLTZ_Iter} and yields a vector mean free path $\vn{\lambda}(\vn{k})=\tau_{\vn{k}}\vn{v}(\vn{k})$ that is collinear with the group velocity. Assuming a vector mean free path of this form in Eq.~\ref{BOLTZ_t_2D_atom} implies a purely odd torkance for the disordered system. In this work, we solve the Boltzmann equation self-consistently, so that the vector mean free path $\mbox{{\boldmath$\lambda$}}(\vn{k})$ is no longer collinear with the group velocity $\vn{v}(\vn{k})$. This property of the vector mean free path is known to generate skew-scattering-induced spin currents \cite{PhysRevLett.104.186403}, but their effect on the torque is not accounted for in our Boltzmann approach, because there are no states propagating out of plane in a film geometry. The absence of skew-scattering-induced spin currents flowing out of plane results in the even torkance being very small compared to the odd torkance in our calculations. We therefore focus on the odd part of the torkance, which is independent of skew-scattering and rather driven by the state dependent relaxation time induced by scattering off impurities.

For all calculations, the electric field is applied along the $x$ direction and magnetization is pointing along the $z$ direction (see Figures~\ref{fig_tork_atoms_FePt_Pt_nogamma} and \ref{fig_tork_atoms_CoCu_nogamma} for system of coordinates). This implies that $t_{xx}$ and $t_{yx}$ are respectively odd and even functions of magnetization direction. We focus on the odd component of the torkance $t_{xx}$ and the corresponding spin accumulation response coefficient $\chi_{yx}$. The C$_{4}$ and C$_{3}$ symmetries, exhibited respectively by the FePt/Pt and Co/Cu films, imply that $\chi_{xy}=-\chi_{yx}$ and $t_{yy}=t_{xx}$. Torkances are given in units of ${\rm ea_{0}}=9.14\times10^{-5}\mu_{B}$T/(V/cm).

The electronic structure of the film is computed using the local density approximation with the parametrization of Vosko, Wilk and Nusair \cite{doi:10.1139/p80-159} and the full-potential relativistic Korringa-Kohn-Rostoker (KKR) method. An angular momentum cut-off of $l_{max} = 3$ was used for the Green functions and the wave functions. The impurity potentials were computed self-consistently in a cluster of atoms including the first and second nearest neighbors using the Jülich KKR impurity-embedding code (KKRimp)~\cite{Bauer:229375}.\\

\subsection{Spin-orbit torques and spin accumulation in an FePt/Pt thin film in the presence of defects}\label{Sec_FePt_Pt}

\subsubsection{Computational details}

We consider two layers of $L{1}_{0}$-FePt oriented along the [001]-direction and terminated with Fe atoms (Fe/Pt/Fe/Pt/Fe stacking sequence) deposited on one side of a six layers Pt(001) film (see Fig.~\ref{fig_tork_atoms_FePt_Pt_nogamma}). Relaxed atomic positions are taken from Ref.~\cite{PhysRevB.91.014417}. The in-plane lattice constant of the film was set to $a= 2.7765$\,\AA. For magnetization out-of-plane, magnetic moments for the Fe-atoms range between 2.93 and 3.06\,$\mu_{B}$, in good agreement with Ref.~\cite{PhysRevB.91.014417} (see Table~\ref{tab_atoms}). The Fermi surface is charted by 24148 $k$-points in the full Brillouin zone~\cite{Bernd_FS}.\\

\subsubsection{Scattering due to defects in FePt (T=0K)}

\begin{figure}
\centering
\includegraphics*[width=8.6cm]{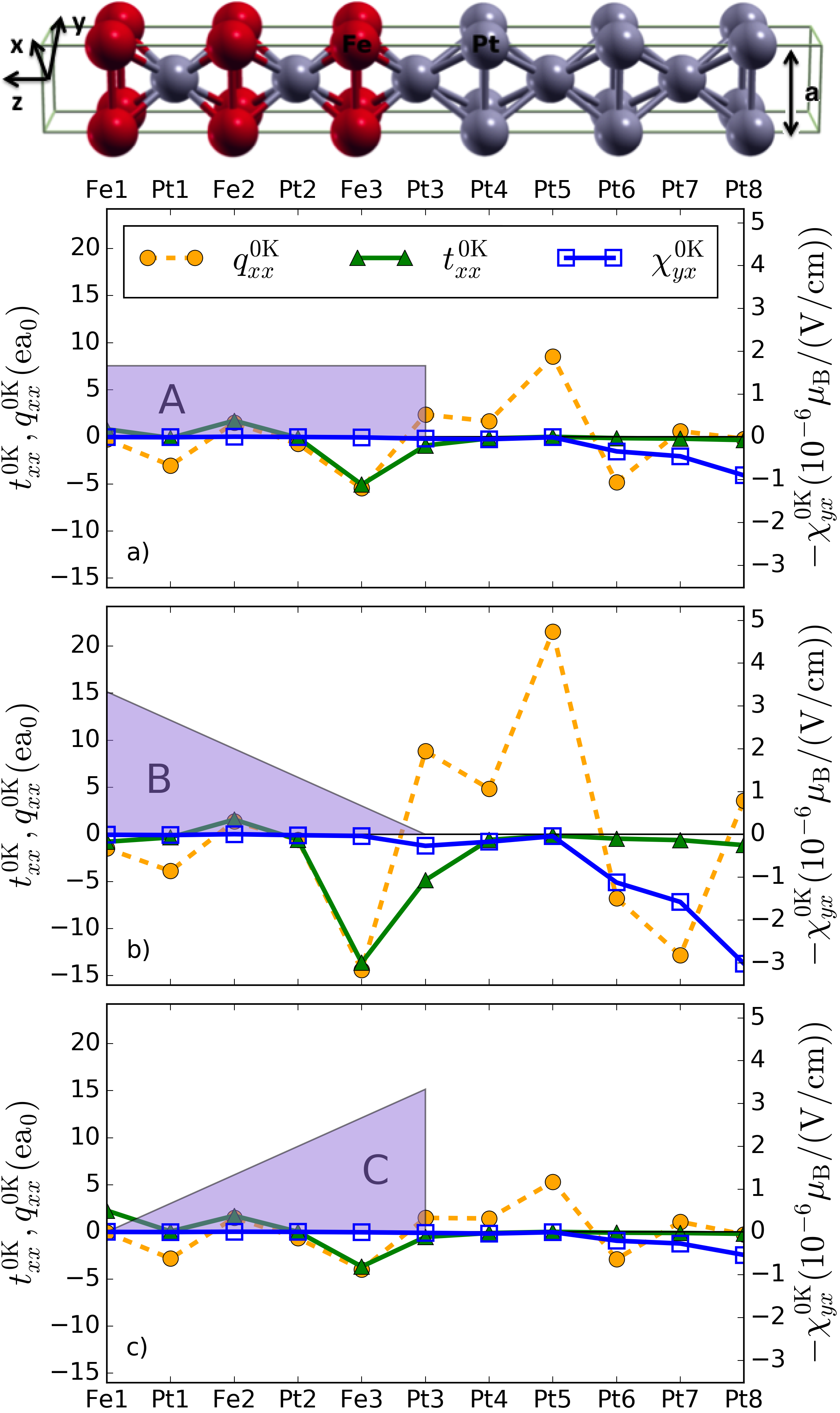}
\caption{\label{fig_tork_atoms_FePt_Pt_nogamma}(top) Illustration of the unit cell of a system of two layers of L1$_0$-FePt oriented along the [001]-direction and terminated with Fe atoms (Fe/Pt/Fe/Pt/Fe stacking sequence) deposited on one side of a six layers Pt(001) film. Fe and Pt atoms are shown in red and gray respectively. Blue shaded areas mark the distribution of defects within the film for cases A, B and C. The concentration of defects per unit cell is equal to $\bar{c}_{imp} = 0.1$  for all three distributions. The scale for zero-temperature torkances (triangles) and spin-flux response coefficients (circles) is shown on the left. The scale for spin-accumulation response coefficents (squares) is given on the right.
}
\end{figure}

Using the transition rates from Eq.~\eqref{transitionrates_gen}, we compute the zero temperature spin accumulation, spin-orbit torques and spin fluxes in an FePt/Pt thin film in the presence of defects in the FePt layer (Fe1 to Pt3 in Fig.~\ref{fig_tork_atoms_FePt_Pt_nogamma}). We shall focus on defects constisting of Fe-atoms being replaced by Pt-atoms and vice versa, which are typically present in epitaxially grown thin films. We consider three different distributions, with defects either distributed homogeneously within the FePt-layer (type A), preferentially at the FePt surface (type B) or at the interface between FePt and Pt (type C), see Fig.~\ref{fig_tork_atoms_FePt_Pt_nogamma}. If not specified otherwise, all calculations are performed for a concentration of 0.1 defects per unit cell. Since the defects are distributed over six atomic layers, this corresponds to a concentration of about 1.7\%.

First, we discuss $\chi^{\rm 0K}_{yx\mu}$, i.e., the response of the spin accumulation to the electric field, Fig~\ref{fig_tork_atoms_FePt_Pt_nogamma}. For all three impurity distributions, the spin accumulation in the magnetic FePt layer is reduced by about one order of magnitude as compared to the Pt side, owing to the competition of spin-orbit fields with the very strong exchange field in FePt. While the spin accumulation in the Pt layers for the distributions A and C is qualitatively similar, its amplitude drastically increases for the distribution B. In that case, the spin accumulation per unit of current reaches sizeable $17.2\times10^{-13} \mu_{B}$/(A/cm$^{2}$) at the bottom of the film, computed from the ratio $\bm{\chi}^{\rm 0K,Pt8}_{yx}/\sigma^{0K}_{xx}$, where the conductivity $\sigma^{0K}_{xx}$ is also computed within the Boltzmann approach outlined above.

\begin{table}[ht!]
\caption{\label{tab_atoms}
Computed spin magnetic moments $\mu_{{\rm at}}$ per atom (in units of $\mu_{B}$) are presented for comparison together with 
spin magnetic moments computed with different method in Ref.~\cite{PhysRevB.91.014417}.
}
\begin{ruledtabular}
\begin{tabular}{ccc}
atomic layer &$\mu_{{\rm at}}$ &$\mu_{{\rm at}}$(Ref.~\cite{PhysRevB.91.014417})\\
\hline
Fe1  &3.060 &3.080\\ 
Pt1  &0.363 &0.403\\
Fe2  &2.934 &3.021\\
Pt2  &0.337 &0.383\\
Fe3  &2.950 &3.040\\
\hline
Pt3  &0.268 &0.297\\
Pt4  &0.038 &0.047\\
Pt5  &0.015 &0.022\\
Pt6  &0.005 &0.009\\
Pt7  &0.006 &0.008\\
Pt8  &0.006 &0.007\\
\end{tabular}
\end{ruledtabular}
\end{table}

\begin{table}[t!]
\caption{\label{effectivefields_FePt_CoCu_nogamma}
Effective magnetic fields $B^{\rm 0K}_y$ for a current density of $j_{x}=10^{7}$A/cm$^{2}$ and torkances $t^{\rm 0K}_{xx}$ for the FePt/Pt film with a concentration of defects per unit cell of $\bar{c}_{\rm imp}=0.1$.
}
\begin{ruledtabular}
\begin{tabular}{ccc}
impurity distribution & $B^{\rm 0K}_y$ (mT) & $t^{\rm 0K}_{xx}$ (ea$_{0}$) \\
\hline
A                     & $-$0.67      &  $-$4.5                     \\
B                     & $-$1.13      & $-$21.5                   \\
C                     & $-$0.14      &  $-$0.7                     \\
\end{tabular}
\end{ruledtabular}
\end{table}

\begin{figure*}[t!]
\centering
\includegraphics*[height=3.05cm]{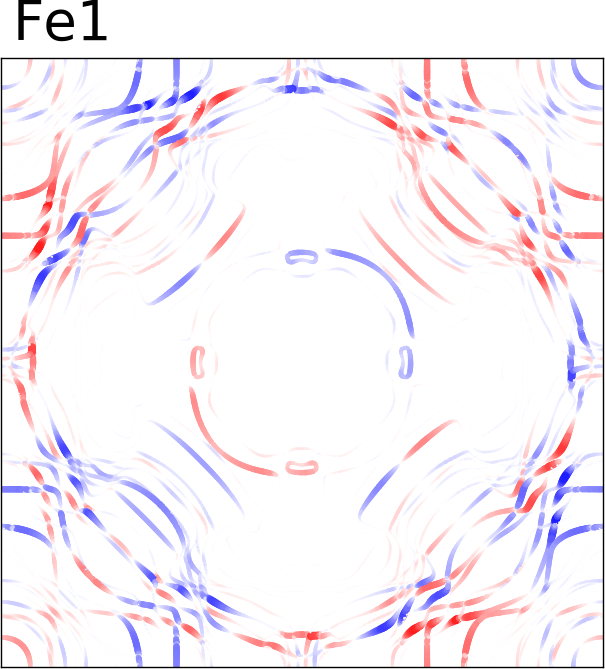}
\includegraphics*[height=3.05cm]{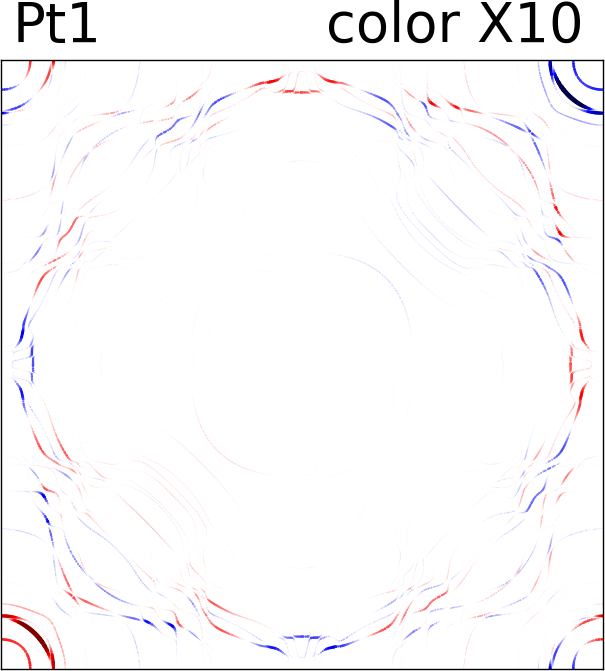}
\includegraphics*[height=3.05cm]{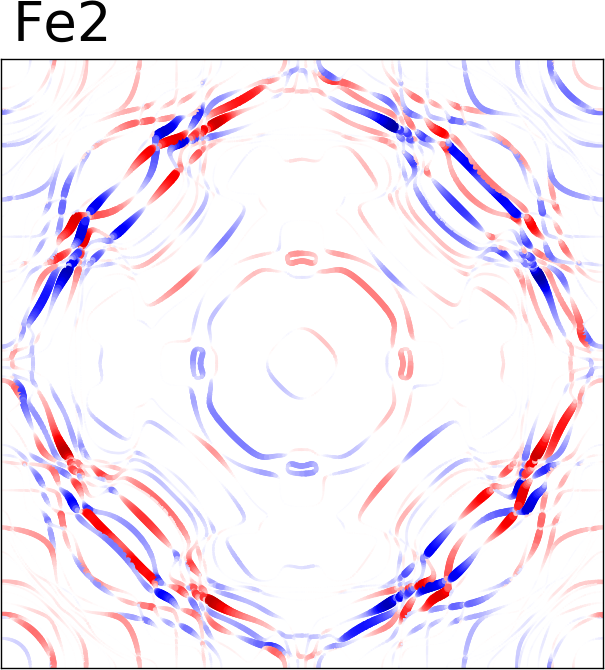}
\includegraphics*[height=3.05cm]{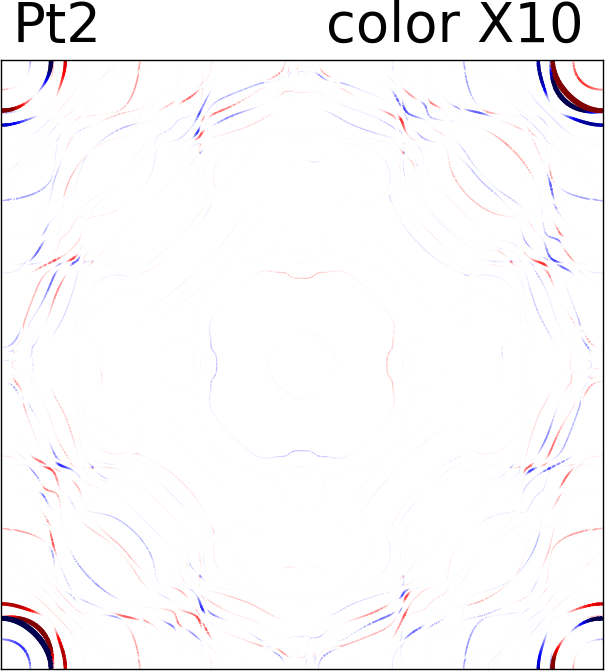}
\includegraphics*[height=3.05cm]{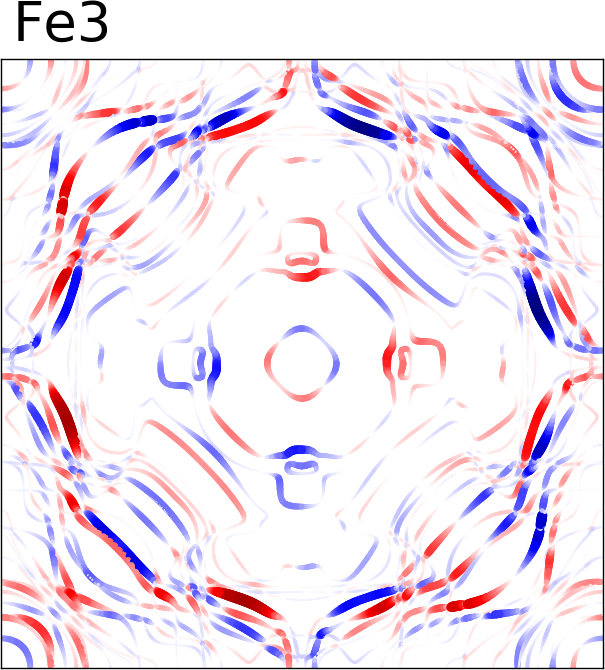}
\includegraphics*[height=3.05cm]{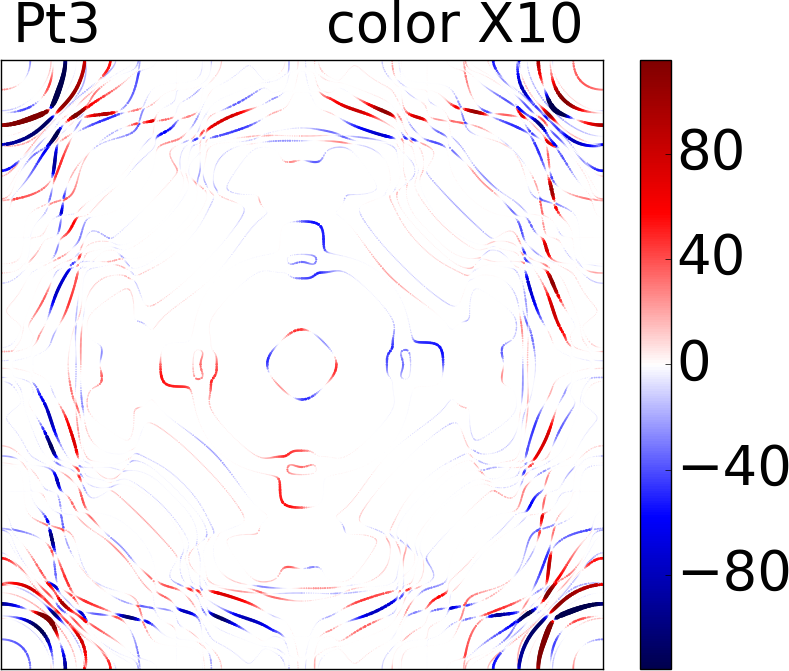}
\caption{\label{torque_FePt_FS}
Expectation values $\langle\mathcal{T}_{x\mu}\rangle$ in meV (color code) for $\mu=$ \{Fe1, Pt1, Fe2, Pt2, Fe3, Pt3\} in the Brillouin zone of the FePt/Pt film. The values are multiplied by a factor of ten for Pt atoms in order to make the details more clear. The thickness of the lines is proportional to the absolute values of $\langle\mathcal{T}_{x\mu}\rangle$ for all plots.}
\end{figure*}

Next, we compute the response of the torque to the electric field, as given by $\vn{t}^{\rm 0K}_{xx\mu}$. According to Fig.~\ref{fig_tork_atoms_FePt_Pt_nogamma}, Fe atoms provide the largest contribution to the torkance, owing to the much larger magnetic moments in these atoms of the order of 3\,$\mu_{B}$, as compared to the induced magnetic moments of the neighboring Pt atoms of the order of 0.3\,$\mu_{B}$ (see Table~\ref{tab_atoms}). This difference in the magnetic moments is directly reflected in the expectation values of the atom-resolved torque operator $\langle\mathcal{T}_{x\mu}\rangle_{\vn{k}}$ computed for the states $\psi_{\vn{k}}$ at the Fermi surface, see Fig.~\ref{torque_FePt_FS}. Notably, Pt3 provides a contribution to the torkance that is comparable in magnitude to that of Fe atoms (Fig.~\ref{fig_tork_atoms_FePt_Pt_nogamma}), eventhough the torque expectation values $\langle\mathcal{T}_{x}^{{\rm Pt3}}\rangle_{\vn{k}}$ for this atom are much smaller (Fig.~\ref{torque_FePt_FS}). In fact, large expectation values of the torque on the FS do not guarantee a large overall torkance because the contributions from different states may cancel. This is best understood if one considers the example of bulk $L{1}_{0}$-FePt, where torque expectation values on the FS would be nonzero, but the response of the torque to an electric field, i.e., the torkance, would vanish as a consequence of inversion symmetry. This explains to a large extent why Pt3 provides a much larger contribution to the torkance than Pt1 and Pt2, whose local environment is almost inversion-symmetric (see Fig.~\ref{fig_tork_atoms_FePt_Pt_nogamma}).

Comparing the atom-resolved torkance to the response of the spin fluxes $\vn{q}^{\rm 0K}_{\mu}$ \cite{ibcsoit}, we find that the torque acting on Fe2 and Fe3 is almost entirely mediated by spin currents, i.e., it comes from the spin-orbit coupling in the Pt-atoms. The large values of the torkance on Fe3 originate primarily from its proximity to the Pt substrate, which provides a large spin current, especially in case B, when the scattering at the interface between two parts of the slab is suppressed, see Fig.~\ref{fig_tork_atoms_FePt_Pt_nogamma}. Even though it tends to be smaller in magnitude, the torque on Fe1 shows a larger deviation from the spin flux. This is a signature of a pronounced breaking of inversion symmetry at the surface responsible for a large torque generated there locally.

In experiments, SOTs are typically measured as effective magnetic fields acting on the magnetization. In order to connect our results to experimentally accessible quantities, we compute the ratio of the effective magnetic field $B^{\rm 0K}_{y}$ to the current density $j^{\rm 0K}_{x}$ using the equation:
\begin{equation}\label{B_eff_field}
\frac{B^{\rm 0K}_y}{j^{\rm 0K}_x}=\frac{1}{M_{\rm {S}}}\frac{t^{\rm 0K}_{xx}}{\sigma^{\rm 0K}_{xx}},
\end{equation}
where $M_{\rm {S}}$ is the magnetic moment per unit cell and $\sigma^{\rm 0K}_{xx}$ is the longitudinal conductivity. This ratio is independent of the total amount of defects when defects are the only source of scattering, since both $t_{xx}$ and $\sigma_{xx}$ scale with the inverse of defect concentration. Taking a typical current density of $10^{7}$A/cm$^{2}$, we find that the effective field at 0K exhibits a variation between $-1.13$ (case B) and $-0.14$ mT (case C), depending on the details of impurity distribution (see Table~\ref{effectivefields_FePt_CoCu_nogamma}). This shows that effective fields at very low temperatures can vary by one order of magnitude depending on fine details of disorder.

The sensitivity of the effective field can be understood very well by looking at the atom-resolved torkances (Fig.~\ref{fig_tork_atoms_FePt_Pt_nogamma}). For distribution A, the torkance is negative and largest on Fe3, while it is smaller and positive  on Fe1 and Fe2 atoms, which results in a negative total torkance of $-$4.5\,ea$_{0}$ (see Table~\ref{effectivefields_FePt_CoCu_nogamma}). For distribution B, the negative torkance on Fe3 is strongly enhanced becoming thus by far the dominant contribution to the very  large total torkance of $-21.5$\,ea$_{0}$. For distribution C, the torkance on Fe3 is noticably
reduced in magnitude as compared to case A, and the competition with enhanced positive torkances on Fe2 and especially Fe1 results in a torkance of only $-0.7$\,ea$_{0}$ in this case.

\begin{figure}[t!]
\centering
\includegraphics*[height=2.8cm]{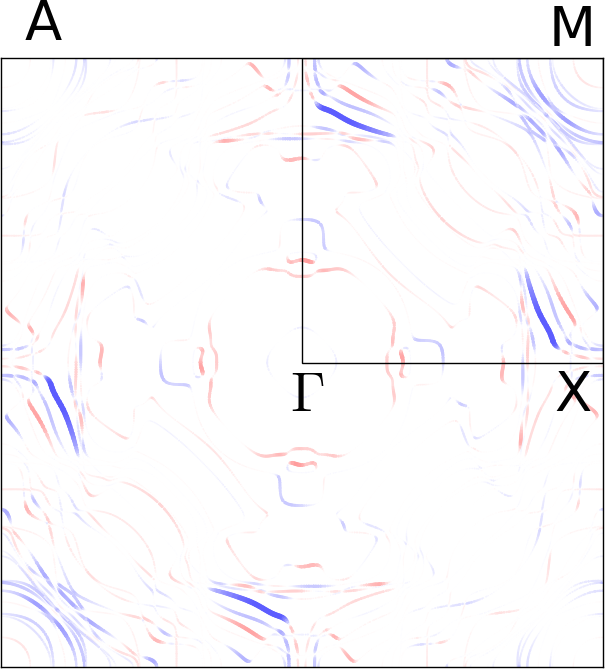}
\includegraphics*[height=2.8cm]{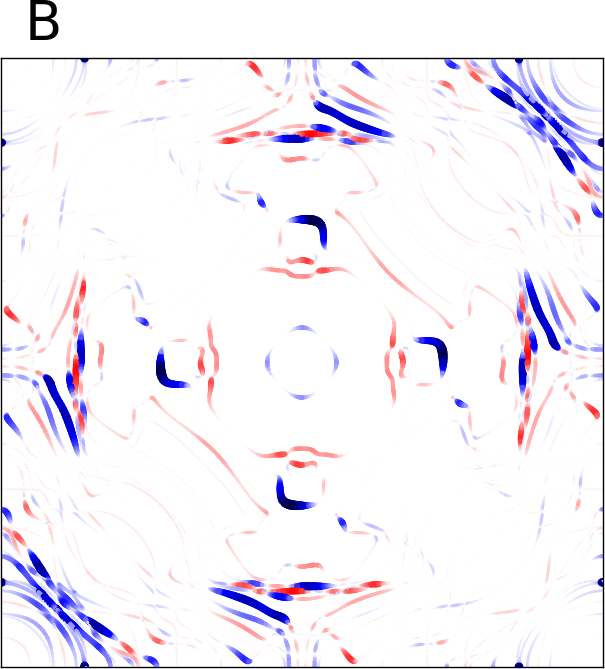}
\includegraphics*[height=2.8cm]{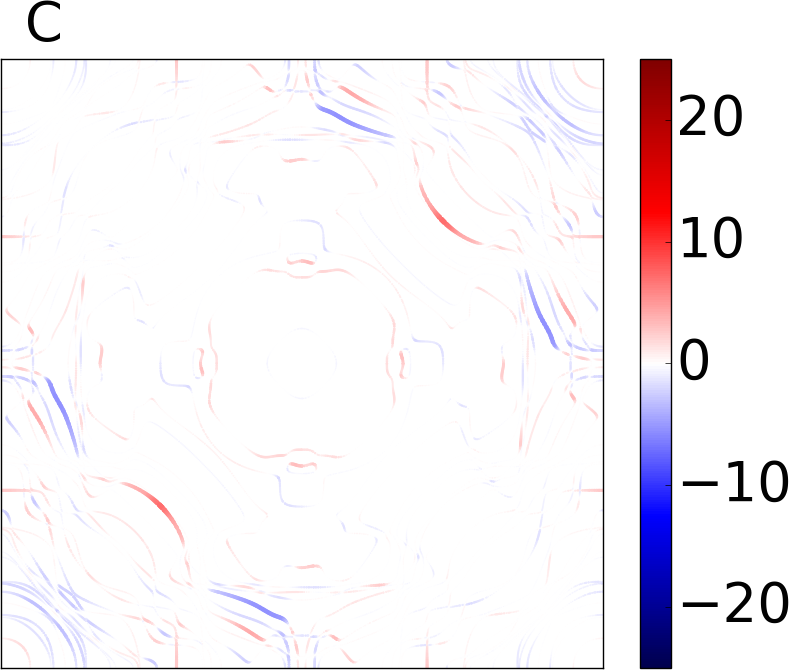}
\caption{\label{torkance_FS}
The total torkance in terms of the Fermi surface distribution of $\langle \mathcal{T}_{x} \rangle_{\vn{k}} \lambda_{x}(\vn{k})/|\vn{v}(\vn{k})|$ [see Eq.~\eqref{BOLTZ_t_2D}], in units of $\hbar$, for an FePt/Pt film with a concentration of defects per unit cell of $\bar{c}_{\rm imp}= 0.1$. Defects are distributed according to scenario A, B and C (see Fig.~\ref{fig_tork_atoms_FePt_Pt_nogamma}). The irreducible Brillouin zone is marked by 
thin black lines.}
\end{figure}

To understand this behavior in more detail, we compute the Fermi surface distribution of the total torkance in terms of the state-resolved quantity $\langle \mathcal{T}_{x} \rangle_{\vn{k}} \lambda_{x}(\vn{k})/|\vn{v}(\vn{k})|$ [see Eq.~\eqref{BOLTZ_t_2D}] for each of the distributions of defects (Fig.~\ref{torkance_FS}). We observe that while the FS distribution of $t_{xx}$ is qualitatively similar in cases A and C, very large negative contributions are visible for case B at the parts of the FS which are suppressed in the case of A and C.  This suggests that the states that are responsible for the increased torkance in case B are localized predominantly at the interface between FePt and Pt parts of the slab, while the presence of finite concentration of impurities at the interface greatly suppresses the contribution of these states in cases A and C. A similar effect can be observed for the torque on Fe1, which is much larger in 
case C with no disorder at the FePt surface, as compared to cases A and B, suggesting that this part of the torque is largely 
driven by the states localized at the surface.

\subsubsection{Defects in FePt at room temperature}

Most SOT experiments are performed at room temperature, when other sources of scattering, such as phonons, play an important role. In the following, we want to understand to which extent the results from the previous section are modified in the presence of other sources of scattering. For this purpose, we use the generalized transition rates from Eq.~\eqref{Pkk_tilde} to compute the spin accumulation response coefficient $\bm{\chi}^{\rm RT}_{\mu}$ and torkance $\vn{t}^{\rm RT}_{\mu}$ in order to simulate the effect of room temperature.

\begin{figure}
\centering
\includegraphics*[width=8.6cm]{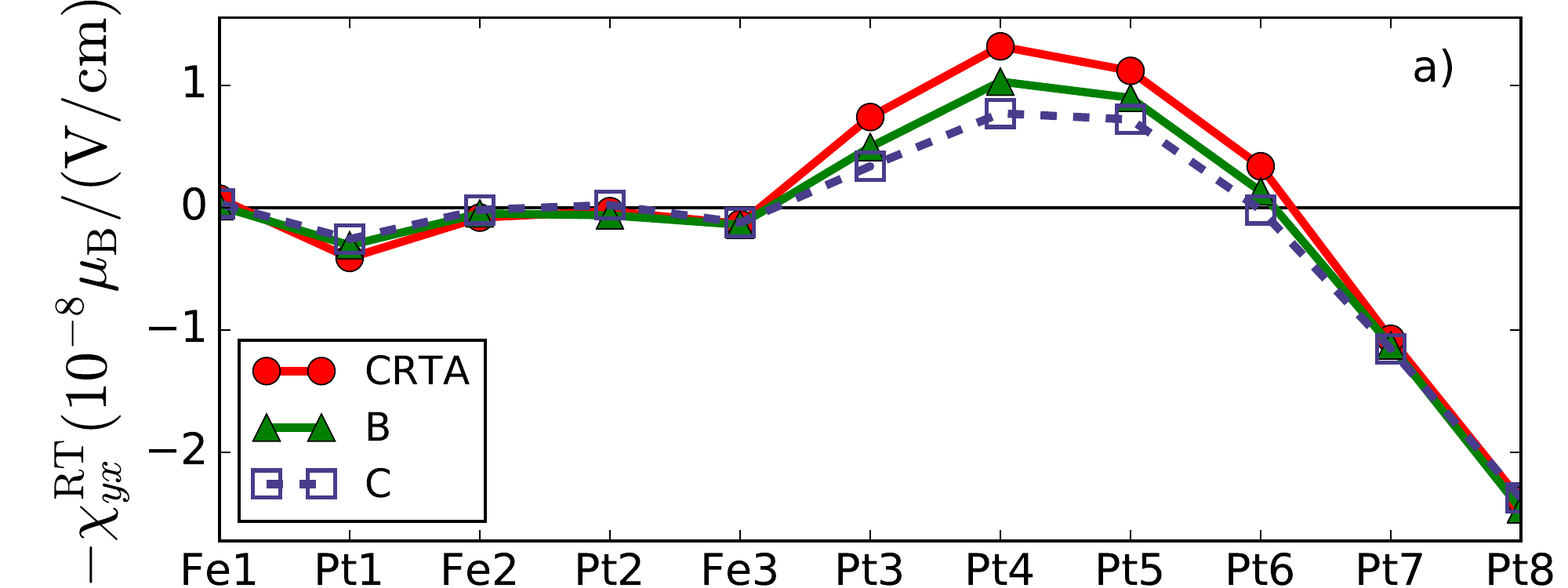}
\includegraphics*[width=8.6cm]{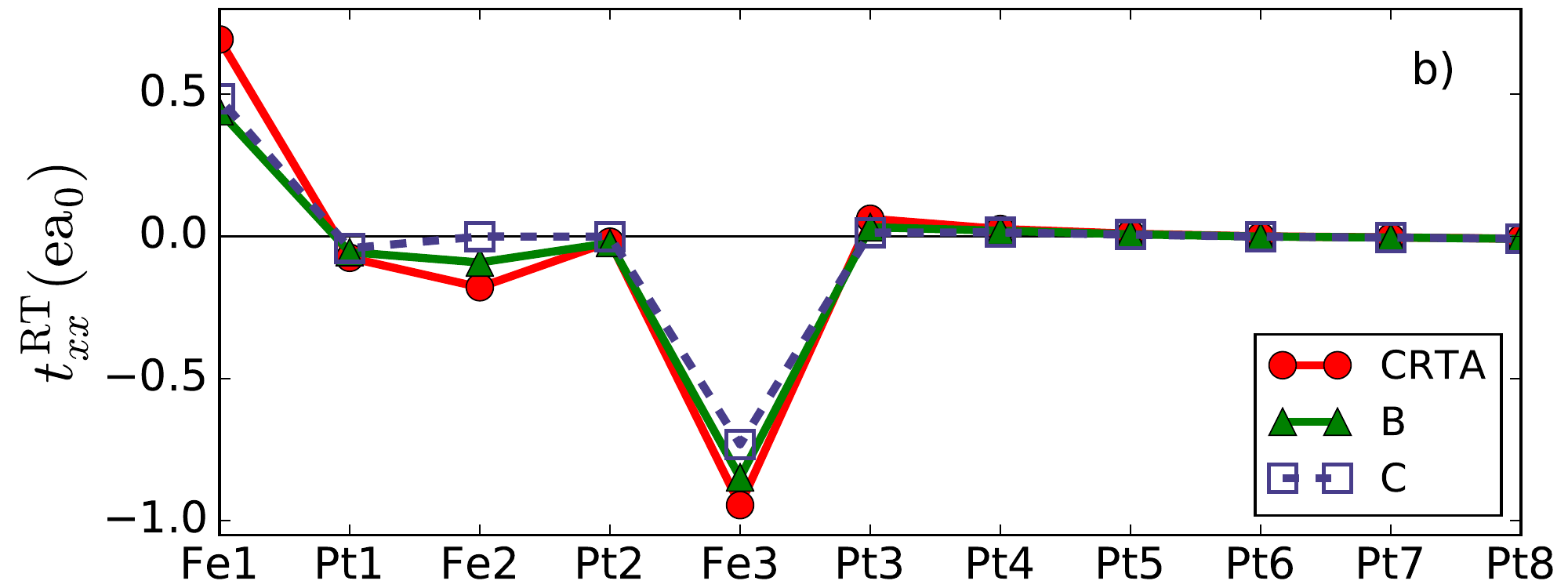}
\caption{\label{fig_tork_atoms_FePt_Pt_noSF}(a) Response coefficient of the spin accumulation, and (b) torkance, computed for three different types of disorder: (squares) constant relaxation time, i.e. $P_{\vn{k}'\vn{k}}$ = 0 in Eq.~\eqref{Pkk_tilde}; (triangles) defects located preferentially at the surface, case B;  (circles) defects located preferentially at the interface, case C. The concentration of defects per unit cell is equal to $\bar{c}_{imp} = 0.1$ for cases B and C. All calculations where performed setting $\Gamma$ = 25 meV in Eq.~\eqref{Pkk_tilde}.}
\end{figure}

As shown in Fig.~\ref{fig_tork_atoms_FePt_Pt_noSF}(a), the magnitude of the spin accumulation as given by $\bm{\chi}^{\rm RT}_{\mu}$ is reduced by about a factor of 100 for case B, and a factor of 25 for case C, as compared to the 0K case, so that at the bottom of the film the spin accumulation per unit of current density is about $4.2\times 10^{-13} \mu_{B}$/(A/cm$^{2}$) for both cases. For case A, values of spin accumulation are intermediate between cases B and C. They are omitted in Fig.~\ref{fig_tork_atoms_FePt_Pt_noSF}(a) for clarity. Overall, the inclusion of temperature clearly washes out the strong difference brought up by different defect distributions at 0K. This observation is consistent with the average value of the relaxation time at zero temperature being such that $\hbar/2\tau^{av}_{\rm 0K}$ is never larger than 8 meV, i.e., considerably smaller than the broadening induced by room temperature. The same trend is observed for the torkance $\vn{t}^{\rm RT}_{\mu}$, shown in Fig.~\ref{fig_tork_atoms_FePt_Pt_noSF}(b), which is decreased by about a factor of 15 for distribution B, and a factor of 5 for distribution C, as compared to the 0K case, so that the large differences between cases B and C at 0K are strongly reduced. The effect of defects is best visible when one compares the calculated spin accumulation and torkance to reference values for a clean system computed using the constant relaxation time approximation (CRTA) with $\Gamma$ = 25 meV and $P_{\vn{k}'\vn{k}}$ = 0 in Eq.~\eqref{Pkk_tilde}, see Fig.~\eqref{fig_tork_atoms_FePt_Pt_noSF}. One clearly sees that overall the impurities tend to reduce the magnitude of the torkance and spin accumulation, with the strongest effect for case C of interfacial disorder.

\begin{figure}
\includegraphics*[width=8.6cm]{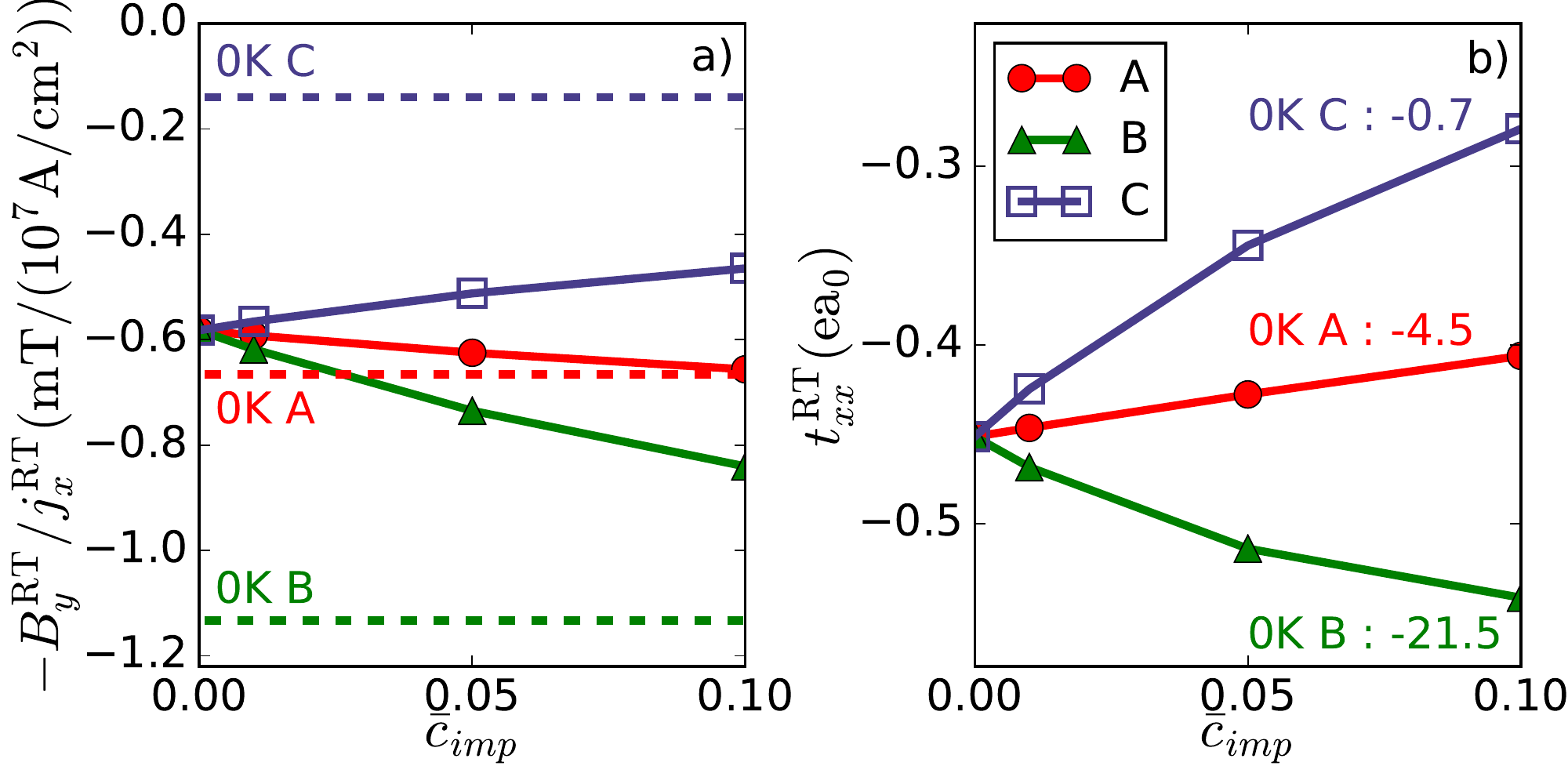}
\caption{\label{fig_B_FePt_Pt_noSF}(a) Effective fields per unit of current density, and (b) total torkances, computed as a function of defect concentration for distributions A, B and C. Horizontal dashed lines and numbers stand for the 0K values, computed with $\bar{c}_{\rm imp}=0.1$ for total torkances. Full lines show the values at room temperature,~i.e., using Eq.~\eqref{Pkk_tilde}.
}
\end{figure}

Although the effect of the impurities on the torque in terms of atomic contributions is rather small, the respective variation of the total torkance and the effective magnetic field can be very large. For case B, comparison to CRTA values in Fig.~\ref{fig_tork_atoms_FePt_Pt_noSF} shows that the positive torque on Fe1 is more strongly reduced than the negative torques on Fe2 and Fe3. This results in a significant increase of the total torkance and the effective field per unit of current density, with respect to CRTA results ($\bar{c}_{imp} = 0$), although the overall magnitude of the torkance remains significantly smaller than that at 0K owing to extreme sensitivity of the current-induced quantities to the amount of disorder in case B, see Fig.~\ref{fig_B_FePt_Pt_noSF}(a-b). For case C, the overall reduction of the negative torques on Fe2 and Fe3 is larger than that of the positive torque on Fe1, Fig.~\ref{fig_tork_atoms_FePt_Pt_noSF}, which results in a drastic decrease of the total torkance and the effective field per unit of current with impurity concentration, Fig.~\ref{fig_B_FePt_Pt_noSF}(a-b). We want to stress that an increase of the torkance when adding impurities (i.e., when reducing the lifetimes of the states on the Fermi surface) might appear counter-intuitive. This effect is a consequence of the torque being a sum of contributions of opposite sign distributed in a non-trivial manner over the Fermi surface. Such behavior of the SOT stands in sharp contrast to that of longitudinal electric conductivity. The effect of the impurities on the $k$-resolved torkance is very hard to see at room temperature, since the fine structure of the SOT distribution observed at zero temperature (Fig.~\ref{torkance_FS}) is washed out by the presence of other sources of scattering as modeled by constant smearing of the states.


\subsection{Spin-orbit torques in a Co/Cu thin film: effect of scattering off impurities}\label{Sec_CoCu}

\subsubsection{Computational details}

We consider a single layer of Co deposited on one side of a six-layer Cu(111) thin film at the fcc-stacking positions (see Fig.~\ref{fig_tork_atoms_CoCu_nogamma}). The in-plane lattice constant of the film was set to $a/\sqrt{2}=2.556$\,\AA, where $a=3.615$\,\AA\phantom{ }is the experimental lattice constant of fcc Cu. The distance between two layers of Cu is set to $a/\sqrt{3}=2.087$\,\AA. The distance between the Co layer and the first Cu layer is set to 1.96\,\AA~\cite{PhysRevB.53.9770}. For the magnetization out-of-plane, the magnetic moment of Co atom is equal to 1.66\,$\mu_{B}$. The Fermi surface is charted by 15810 $k$-points in the full Brillouin zone~\cite{Bernd_FS}.\\

\subsubsection{Scattering due to impurities (T=0K)}

\begin{figure}
\centering
\includegraphics*[width=8.6cm]{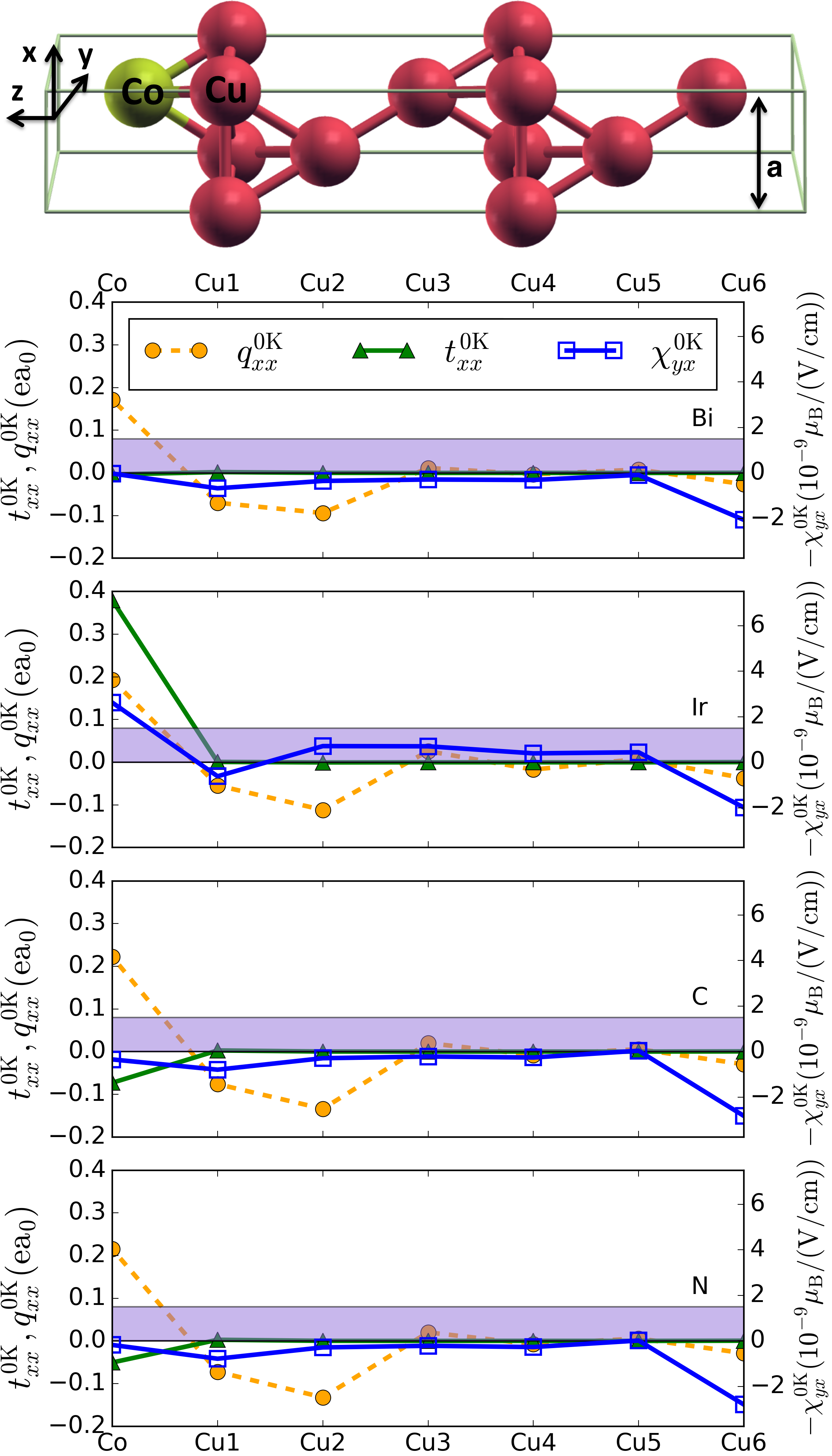}
\caption{\label{fig_tork_atoms_CoCu_nogamma}(top) Sketch of the unit cell of one layer of Co deposited on six layers of Cu(111). Co and Cu atoms are shown in yellow and red respectively. Blue shaded areas mark the homogeneous distributions of Bi, Ir, C and N impurities. The concentration of impurities per unit cell is equal to $\bar{c}_{imp} = 0.1$ for all distributions. The scale for zero temperature torkance (triangles) and spin-flux response coefficient (circles) is shown on the left. The scale for the spin-accumulation response coefficent (squares) is given on the right.
}
\end{figure}

We compute the zero-temperature spin accumulation, spin-orbit torques and spin fluxes in a Co/Cu film in the presence of Bi, Ir, C and N impurities. The impurities are distributed homogeneously in  Co and six Cu layers of the film, see Fig.~\ref{fig_tork_atoms_CoCu_nogamma}. If not specified otherwise, all our calculations are performed for a concentration of 0.1 impurities per unit cell. Since the impurities are distributed over seven atomic layers, this corresponds to a concentration of impurities of about 1.4\%.
\begin{table}[t!]
\caption{\label{effectivefields_CoCu_nogamma}
Effective magnetic fields $B^{\rm 0K}_y$ for a current density of $j_{x}=10^{7}$A/cm$^{2}$ and torkances $t^{\rm 0K}_{xx}$ for a Co/Cu film with a concentration of impurities per unit cell of $\bar{c}_{\rm imp}=0.1$.}
\begin{ruledtabular}
\begin{tabular}{ccc}
Impurity type & $ B^{\rm 0K}_y$ (mT) & $t^{\rm 0K}_{xx}$ (ea$_{0}$) \\
\hline
Bi                     &   0.00     &   0.00        \\
Ir                      &   1.61     &   0.38         \\
C                      & $-$0.33     &  $-$0.07         \\
N                      & $-$0.22     &  $-$0.05          \\
\end{tabular}
\end{ruledtabular}
\end{table}

As visible in Fig.~\ref{fig_tork_atoms_CoCu_nogamma}, the spin-accumulation coefficient $\chi^{\rm 0K}_{yx,\mu}$ at the bottom of the Co/Cu film is three orders of magnitude smaller than in the case
of the FePt/Pt film studied in Section \ref{Sec_FePt_Pt}, which is due to the much smaller spin-orbit coupling strength of Cu, and finite concentration of impurities at the bottom of the Co/Cu film. In the case of Bi, C and N impurities, the spin accumulation is positive almost everywhere in the Cu substrate and has relatively small values in the Co atoms: $0.00$, $0.34$ and $0.18\times10^{-9}\mu_{\rm B}$/(V/cm) for Bi, C and N impurities, respectively. The overall situation is quite different for Ir impurities, where the spin accumulation is negative for layers from Cu2 to Cu5, and it reaches as much as $-2.61\times10^{-9}\mu_{\rm B}$/(V/cm) on the Co atoms. 

\begin{figure*}
\includegraphics[height=3.6cm]{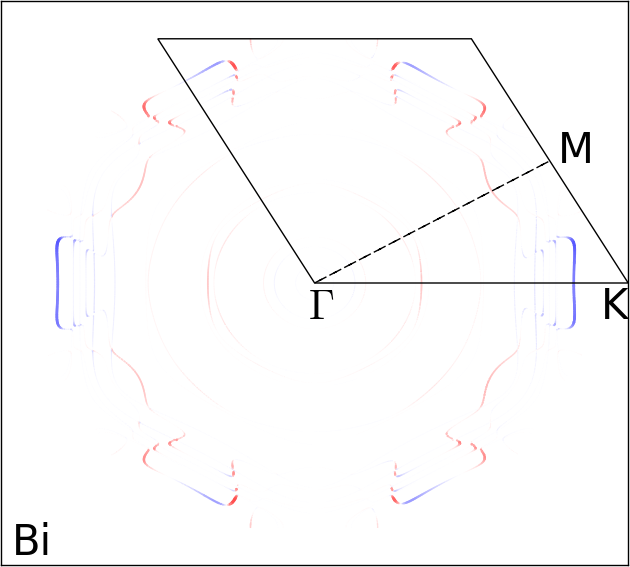}
\includegraphics[height=3.6cm]{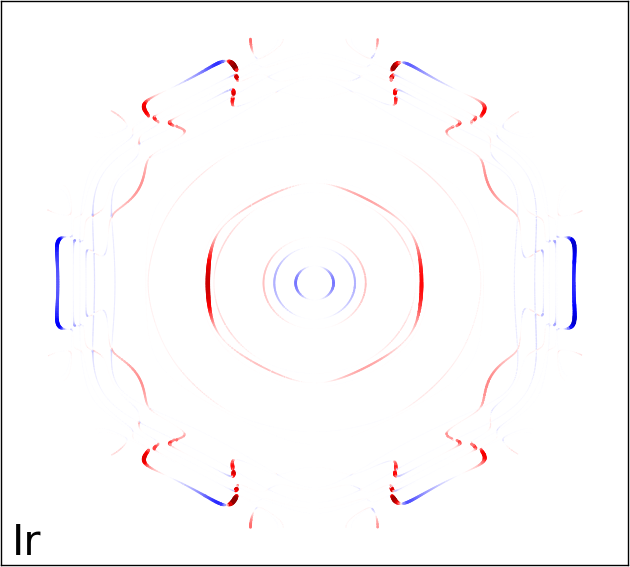}
\includegraphics[height=3.6cm]{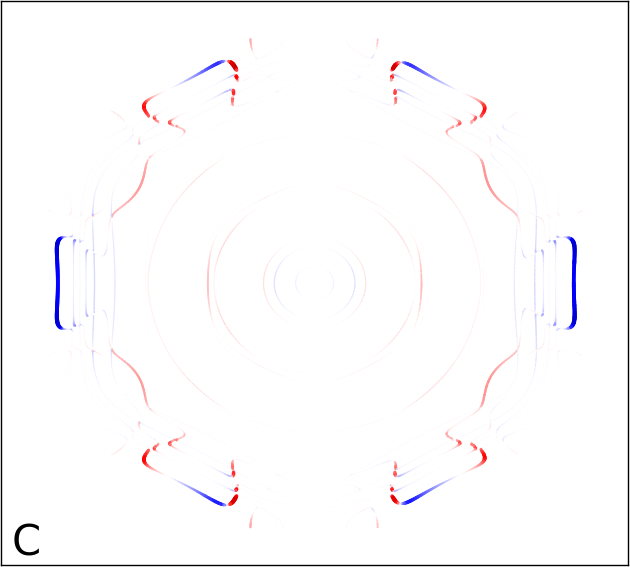}
\includegraphics[height=3.6cm]{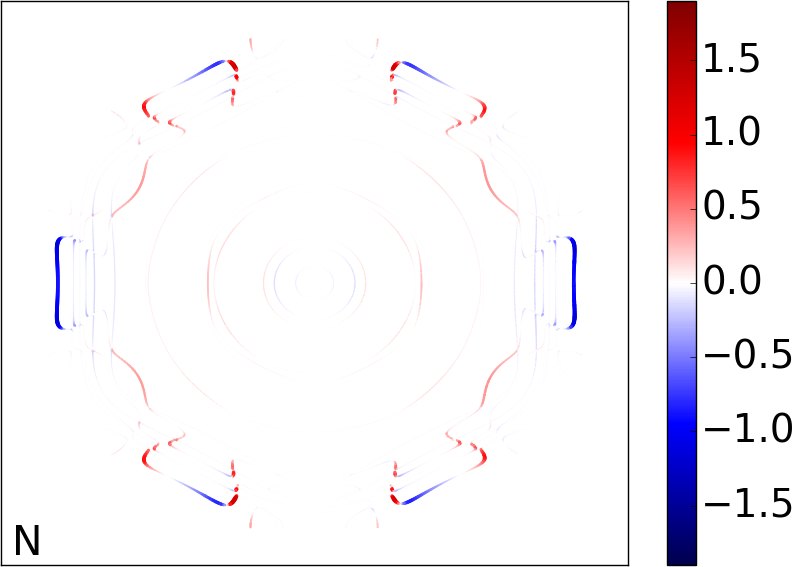}
\caption{\label{torkance_FS_CoCu_Bi}
The total torkance in terms of the Fermi surface distribution of $\langle \mathcal{T}_{x} \rangle_{\vn{k}} \lambda_{x}(\vn{k})/|\vn{v}(\vn{k})|$ [see Eq.~\eqref{BOLTZ_t_2D}], in units of $\hbar$, for a Co/Cu film with a concentration of Bi, Ir, C or N impurities per unit cell of $\bar{c}_{imp} = 0.1$. The irreducible Brillouin zone is shown by solid lines and the dashed line is a guide to the eyes.
}
\end{figure*}

The torkance $t^{\rm 0K}_{xx,\mu}$ on the Co atoms follows the general trend of the spin accumulation: it reaches about 0.4\,ea$_{0}$ in the case of Ir-doping, and only $-$0.1\,ea$_{0}$ for the case of light C and N impurities, while it completely vanishes for the film doped with Bi impurities. We observe that, in contrast to the FePt/Pt system, the torkance has significant magnitude only within the layer of Co atoms, i.e., the torque exerted on the very small induced magnetization of the Cu1 atom ($\approx 0.004 \mu_{B}$) is negligible. On the other hand, it is clear that the case of Co/Cu film presents an example of a system where the torque on the ferromagnet is not driven by the absorbed spin flux: the spin flux $q^{\rm 0K}_{xx,\mu}$ and the torque on the Co atoms, although generally having similar magnitudes, have no visible correlation. Namely, as opposed to the 
behavior of the torque, the spin flux into the Co atoms is almost independent of the type of impurity, and its value remains close to $0.2$ ea$_{0}$. 
The discrepancy between the spin flux
and the torque on Co can be attributed to the effect of the local spin-orbit coupling in the Co atoms \cite{PhysRevLett.105.126602}.

The total torkances $t^{\rm 0K}_{xx}$ and the effective fields per unit of current density $B^{\rm 0K}_{y}$ are given in Table~\ref{effectivefields_CoCu_nogamma}. The strong dependence of these values on the impurity type can be understood from careful inspection of the FS distribution of the torkance (Fig.~\ref{torkance_FS_CoCu_Bi}). In the case of C and N impurities, the largest contribution to the torkance comes from the outer part of the Brillouin zone, where the torkance is mostly negative. The dominant
contribution of these FS parts yields total torkances of respectively $-0.07$\,ea$_{0}$ (C) and $-0.05$\,ea$_{0}$ (N). In the case of Bi impurities, the contribution of the outer parts of the FS  is suppressed, indicating that these states scatter much stronger off heavy Bi impurities, than off lighter C and N impurities. The overall picture is quite different for Ir impurities, where a FS loop enclosing the $\Gamma$-point yields an increased positive contribution to the torkance. This indicates that the electrons in the latter states are much less effected by scattering off Ir impurities, than off Bi, C or N impurities. It results in a torkance of 0.38~ea$_{0}$, which is of an opposite sign and a much larger magnitude as compared to other impurity types. For the electronic states of the FS loop enclosing the $\Gamma$-point, the much stronger scattering induced by Bi as compared to Ir impurities is suprising given the similar atomic numbers of these elements. The observation that Ir and Co have similar valence shells, respectively 5d$^{7}$6s$^{2}$ and 3d$^{7}$4s$^{2}$, suggests that the scattering induced by Ir impurities in the Co layer should be rather weak. Since the states of this FS loop are predominantly localized on the Co layer, 
%at $\sim90\%$, 
the similarity of the valence shell of the impurity with the one of the Co atoms is crucial. Overall, we can conclude that at low temperatures the total torque on the magnetization can depend very sensitively on the impurity type used to dope
the substrate, both in sign and in magnitude. This dependence provides a powerful tool to control the SOT properties in magnetic heterostructures.

\subsubsection{Impurities at room temperature}

\begin{figure}
\includegraphics[width=8.6cm]{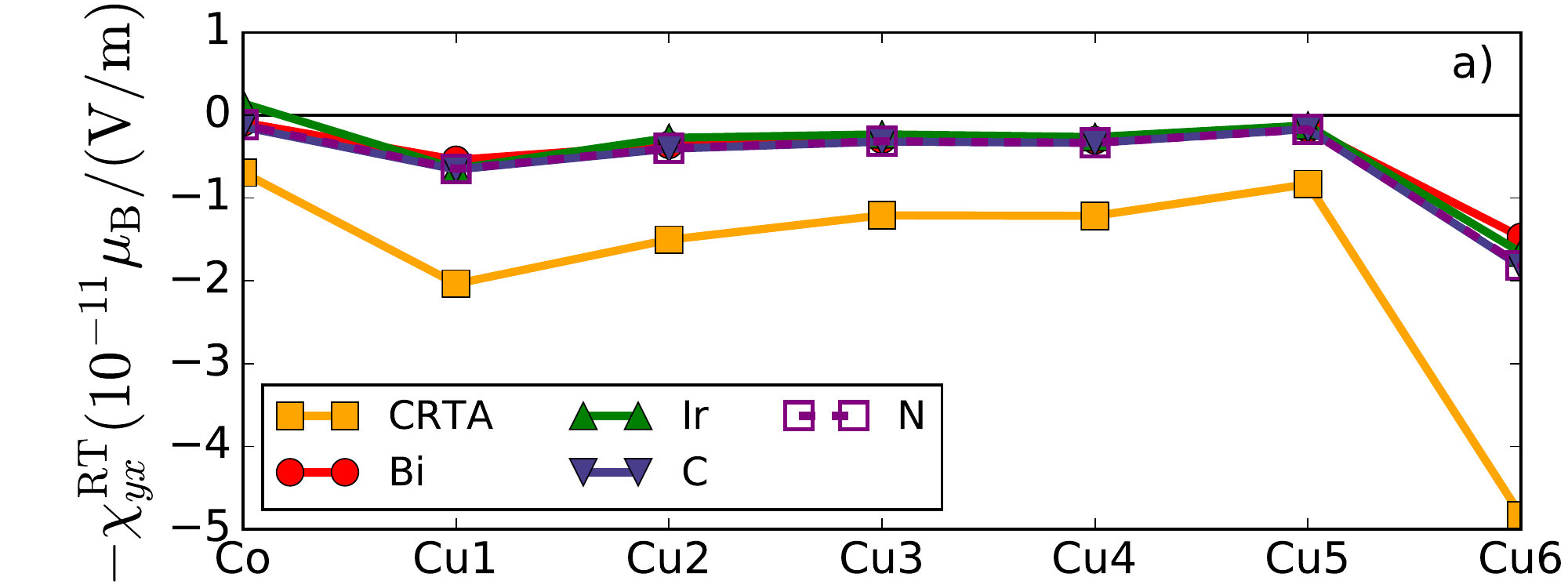}
\includegraphics[width=8.6cm]{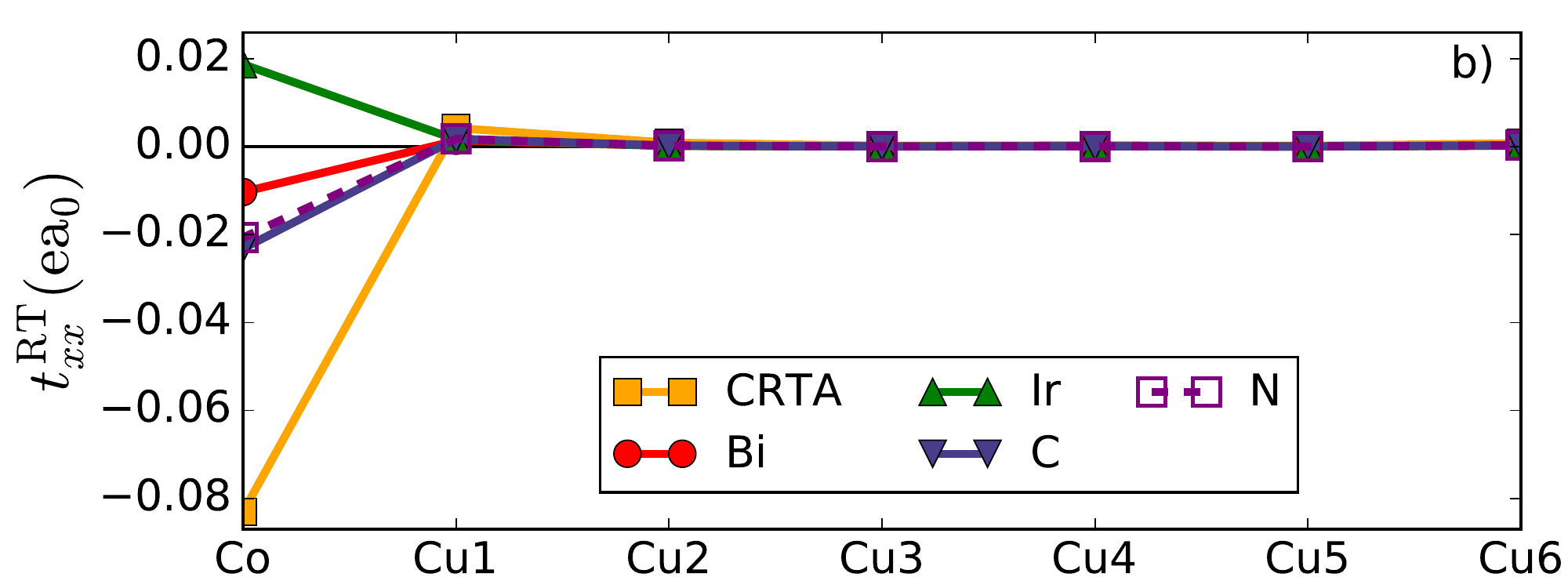}
\caption{\label{fig_sa_atoms_Co_Cu} (a) Response coefficient of the spin accumulation, and (b) torkance, in a Co/Cu film for the clean system in CRTA (full squares), and in the presence of Bi (circles), Ir (triangles up), C (triangles down) and N (open squares) impurities. The concentration of impurities per unit cell is equal to $\bar{c}_{imp} = 0.1$ for all impurity types. All calculations where performed with $\Gamma$ = 25 meV in Eq.~\eqref{Pkk_tilde}.
}
\end{figure}

\begin{figure}
\includegraphics[width=8.6cm]{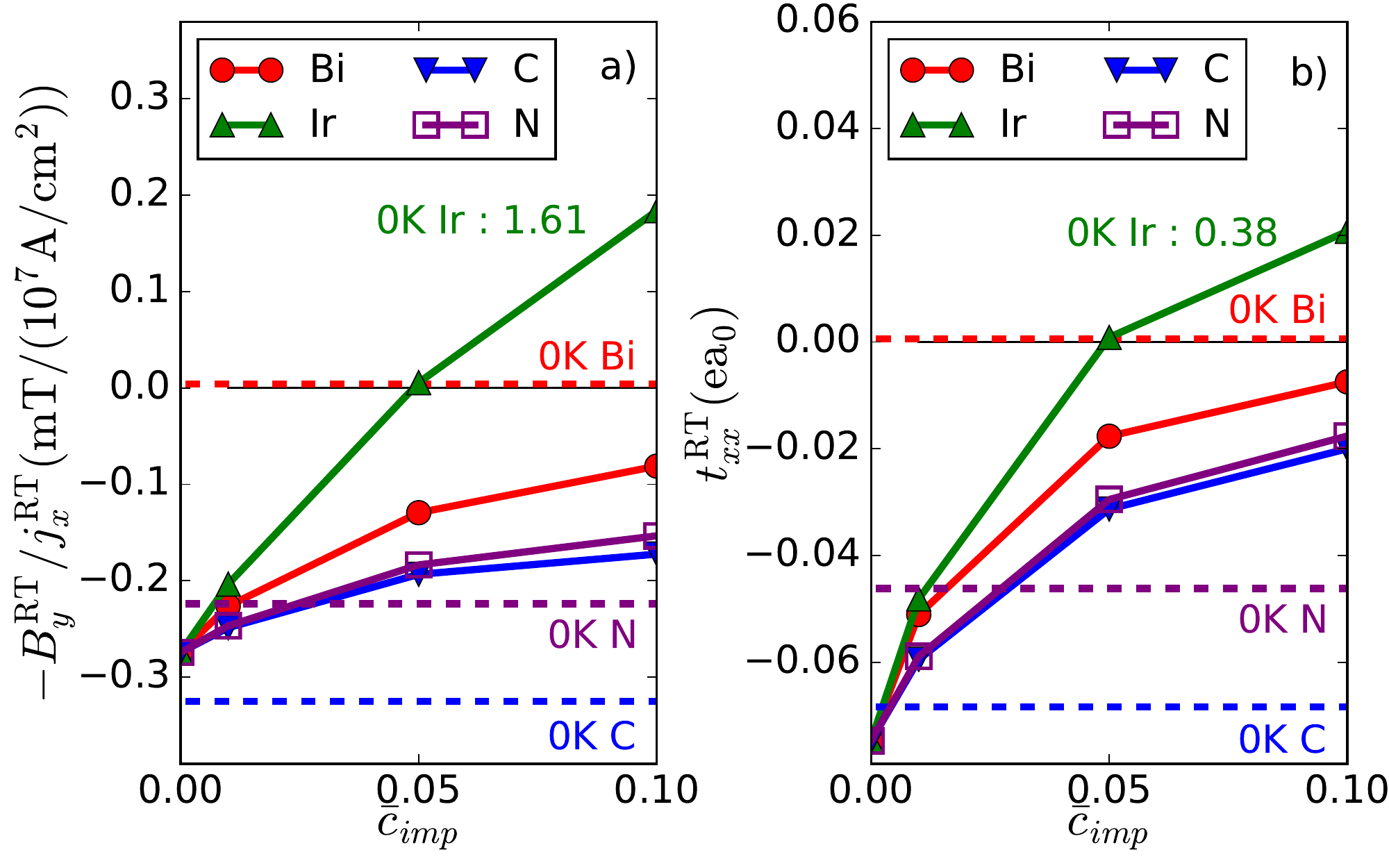}
\caption{\label{fig_B_Co_Cu}(a) Effective fields per unit of current density, and (b) total torkances, computed as a function of concentrations of Bi, Ir, C and N impurities. Horizontal dashed lines and numbers stand for the 0K values, computed with $\bar{c}_{\rm imp}=0.1$ for total torkances. Full lines show the values at room temperature,~i.e., using Eq.~\eqref{Pkk_tilde}.
}
\end{figure}

In the following, we investigate how the state-dependent scattering off impurities influences the spin accumulation and the SOT when other sources of scattering are present, as it is the case at room temperature. For this purpose, in analogy to the preceeding section, we use the generalized transition rates from Eq.~\eqref{Pkk_tilde} to compute the room temperature spin accumulation response coefficient $\chi^{\rm RT}_{yx,\mu}$ and torkance $t^{\rm RT}_{xx,\mu}$, presenting the results of the calculations in Fig.~\ref{fig_sa_atoms_Co_Cu} and Fig.~\ref{fig_B_Co_Cu}.

As shown in Fig.~\ref{fig_sa_atoms_Co_Cu}a, the amplitude of the spin accumulation along the Co/Cu film is reduced by about two orders of magnitude as compared to the zero temperature values. The sign of the spin accumulation in the Co layer is consistent with the zero temperature case, i.e., it is negative for Ir impurities (note that the spin accumulation at Co surface for this case is negligible as compared to the corresponding value at the Cu-side of the slab, in contrast to the 0K results) and positive for Bi, C and N impurities. The same is true for the torkance on the Co-atom, which is positive only in the case of Ir-impurities, see Fig.~\ref{fig_sa_atoms_Co_Cu}b. However, the large variation of the torkance with the type of impurity is strongly reduced from 0.45~ea$_{0}$ at zero temperature to 0.04~ea$_{0}$ at room temperature. 

The effect of defects is best visible if one compares the calculated spin accumulation and torkances to a reference calculation for a clean system. For this purpose, we compute also the spin accumulation and the torkance within the constant relaxation time approximation (CRTA), setting $\Gamma$ = 25 meV and $P_{\vn{k}'\vn{k}}$ = 0 in Eq.~\ref{Pkk_tilde}, see Fig.~\ref{fig_sa_atoms_Co_Cu}. We find that the presence of the impurities reduces the torkance by about a factor of four as compared to the CRTA values. These results indicate that, although the variation of the local RT spin accumulation and torque as a function of the impurity type is qualitatively well-reproduced by the zero temperature values, the overall magnitude of the latter quantities is extremely sensitive to the fine details of the disorder for this system. This stands somewhat in contrast to the situation we encountered for FePt/Pt film when considering the effect of 
disorder distribution in the FePt layer, Fig.~\ref{fig_tork_atoms_FePt_Pt_noSF}.

We conclude our analysis by plotting in Fig.~\ref{fig_B_Co_Cu} the room temperature total torkances and effective magnetic fields per unit of current, as a function of impurity concentration. We find that the torque exerted on the magnetization of the Co/Cu film is affected by the presence of the impurities in two ways. On the one hand, there is a tendency for the torkance to decrease as a function of the impurity concentration, because the average relaxation time is reduced. For a concentration of impurities per unit cell of $\bar{c}_{\rm imp}=0.1$, this results in the torkance being reduced by about a factor of 10 for Bi impurities and 4 for C and N impurities. On the other hand, the state-dependent scattering mediated by impurities tends to push the torkance toward the zero-temperature impurity-type dependent values. The competition between these two trends explains the characteristic behavior of the torkance with increasing impurity concentration, in particular, the change of sign observed for Ir impurities at the concentration of $\bar{c}_{\rm imp}=0.05$, and the eventual splitting of the curves for Bi, C and N impurities. Overall, the range of values displayed in Fig.~\ref{fig_B_Co_Cu} clearly demonstrates that
the combination of proper impurity type with proper concentration allows us to engineer the desired properties
of the disorder-driven SOT in magnetic films.

\section{Conclusions}\label{Conclusions}
Using the \textit{ab initio} Boltzmann formalism, we computed the current-induced SOT in $L{1}_{0}$-FePt/Pt and Co/Cu thin films in the presence of impurities. The transition rates between electronic states were obtained by
considering microscopic scattering off isolated impurities as treated from first principles. Our approach goes significantly beyond the standard description of extrinsic SOT, which is usually based on the constant relaxation-time approximation. In $L{1}_{0}$-FePt/Pt thin films, we demonstrate the crucial dependence of the SOT on the distribution of impurities in the FePt layers, which we explain by a cancellation of opposite contributions at the Fermi surface to the total torque in the system. Moreover, we predict a large spin accumulation in the Pt layers and show that a large part of the SOT is mediated by spin currents. Taking Co/Cu thin films as an example, we show the crucial dependence of the sign and magnitude of the SOT on the type of impurities, and provide evidence that the sign of the spin-orbit torque on the magnetization can be tuned by the concentration of impurities. We explain the sensitivity of the torque to the impurity type by the state-dependent relaxation-time induced by the scattering off impurities. 

\section{Acknowledgements}

We gratefully acknowledge computing time on the supercomputers of J\"ulich Supercomputing Center and RWTH Aachen University, as well as the funding under SPP 1538 “Spin Caloric Transport” and SPP 1666 “Topological Insulators” of the Deutsche Forschungsgemeinschaft. This publication is part of a project that has received funding from the European Union’s Horizon 2020 research and innovation programme under grant agreement No 665095.
\bibliography{letter}

\end{document}